\documentclass[a4paper,10pt,onecolumn,nofootinbib,aps,prd]{revtex4-2}
\usepackage[latin1]{inputenc}
\usepackage{amsmath}
\usepackage{amsfonts}
\usepackage{amssymb,amsthm}
\usepackage{graphicx}
\usepackage[left=2cm, right=2cm, top=2.5 cm, bottom=2.5 cm]{geometry}
\usepackage{float}
\usepackage{tikz}
\usetikzlibrary{shapes,arrows,shadings,shadows}
\usepackage{graphicx}
\usepackage[compat=1.1.0]{tikz-feynman} 
\usepackage{hyperref}
\usepackage[english]{babel}
\usepackage{subcaption}
\usepackage[justification=centering,singlelinecheck=false]{subcaption}
\usepackage[justification=RaggedRight,singlelinecheck=false]{caption}
\usepackage{amsfonts}
\usepackage{braket}
\usepackage{mathrsfs}
\usepackage{empheq}
\usepackage[shortlabels]{enumitem}
\usepackage{pgfplots}
\usepackage{bigints}
\usepackage{tikz, pgf}
\usetikzlibrary{decorations.pathmorphing}
\usetikzlibrary{arrows.meta}
\tikzset{%
  >={Latex[width=2mm,length=2mm]},
            base/.style = {rectangle, rounded corners, draw=black,
                           minimum width=4cm, minimum height=1cm,
                           text centered, font=\sffamily},
}
\usetikzlibrary{calc,bending,decorations.markings}
\usepackage{setspace}

\usepackage{graphicx}
\usepackage{dcolumn}
\usepackage{bm}

\begin{document}

\title{Some new observations for the Georgi-Machacek scenario with triplet Higgs Scalars}

\author{Rituparna Ghosh}
\email{rg20rs072@iiserkol.ac.in}
\author{Biswarup Mukhopadhyaya}%
\email{biswarup@iiserkol.ac.in}

\affiliation{Department of Physical Sciences, Indian Institute of Science Education and Research, Kolkata, Mohanpur - 741246, India}%

\begin{abstract}
The Georgi-Machacek model, introducing a complex  and a real scalar triplet
as additional components of the electroweak symmetry breaking sector, enables substantial triplet contributions to the weak gauge boson masses, subject to the equality of the complex and the real triplet vacuum expectation values (vev) via a custodial SU(2) symmetry.  We present an updated set of constraints on this scenario, from collider data (including those from 137/139~fb$^{-1}$ of luminosity at the Large Hadron Collider), available data on the 125-GeV scalar,  indirect limits and also theoretical restrictions from vacuum stability and unitarity. It is found that some bounds get relaxed, and the phenomenological potential of the scenario is more diverse, if the doubly charged scalar in the spectrum can decay not only  into  two like-sign $W$'s but also into one or two singly charged scalars. Other interesting features are noticed in a general approach, such as substantial $\gamma\gamma$ and $Z\gamma$ branching ratios of the additional custodial singlet scalar, and appreciable strength of the trilinear interaction of a charged scalar, the $W$ and the $Z$. Finally, we take into account the possibility of custodial  SU(2) breaking, resulting in inequality of the real and the complex scalar vevs which too in principle may allow large triplet contribution to weak boson masses. Illustrative numerical results on the modified limits and
predictions are presented, once more taking into account all the constraints
mentioned above.
\end{abstract}

\maketitle
\newpage
\section{Introduction}
\indent As the electroweak symmetry breaking sector continues to be closely scrutinised both theoretically and experimentally, a query persists far and wide. It is: can the
vacuum expectation values (vev) of other 
scalars, not necessarily $SU(2)_L$ doublets, also contribute substantially to the $W$-and $Z$-boson masses, besides the 125-GeV particle that arises overwhelmingly out of a doublet?\cite{125s1} SU(2) triplet scalars are especially interesting in this connection, since  (a) they may occur in some Grand Unified Theories (GUT)\cite{GUT2,GUT1} as well as in left-right symmetric scenarios\cite{LR3,LR1,HHG,LR2} and (b)they offer a mechanism
for generating Majorana masses for left-handed neutrinos, called the Type-II seesaw mechanism\cite{seesaw1}\cite{seesaw2}. However, there is a strong constraint on, say, a single Y=2 complex triplet vev from the $\rho$-parameter\cite{HHG}\cite{obli1}, whose tree-level value is given by $\rho = {\frac{m^2_W}{m^2_Z \cos^2 \theta_W}} = 1$, where $\theta_W$ is the weak boson mixing angle. The vev of a standalone triplet thus cannot exceed about $4 GeV$, hence its contribution to the weak boson masses is rather meagre. A frequently discussed model in this context is the one proposed first by Georgi and Machacek (GM), where one complex ($\chi$, Y=2) and one real ($\xi$,Y=0) triplet were introduced in addition to the doublet $\Phi$ of the standard model (SM)\cite{GM}\cite{CG}. Such a model can be associated with `composite Higgs' scenarios\cite{ch}, but it is of sufficient interest on its own. It ensures $\rho = 1$ at tree-level if the two triplets have equal vev, ensured
with the help of a global SU(2) as custodial symmetry\cite{cust}\footnote{Studies on variants of the model including additional symmetries are also found in the literature, but we are not restricting ourselves by such considerations here.\cite{Limaz2}}. Thus, in the simplest case, one has $v_\chi = v_\xi$. In this case, $s_{H} = \frac{2\sqrt{2}v_{\chi}}{\sqrt{v^{2}_{\Phi} + 8v^{2}_{\chi}}}$ emerges
as a measure of the triplet contribution to the gauge boson masses. Attempts have been made in recent works to obtain upper bounds on this important parameter from experimental as well as theoretical considerations, mostly as  functions of the mass of the doubly-charged component of the triplet  $\chi$, which is one distinct ingredient of this scenario. The available  limits used data from the Large Hadron Collider (LHC) as well as indirect limits such as those from rare decay processes as also precision electroweak measurements.

In this work, we generalise and update these limits, which enable one to extend the
region of the GM parameter space that can be constrained, and also include additional
possibilities in the particle spectrum of this kind of a theory. In particular, the new features of our analysis are as follows:

\begin{enumerate}
\item Available updated limits from the LHC data have been incorporated. Most important of these are the new limits on doubly charged scalar production via vector boson fusion (VBF)\cite{VBF2} and also Drell-Yan (DY)\cite{DY} processes. The latter is particularly
important, because it predicts production cross-sections which are four times as much as
those for singly-charged scalars.   

\item The previous search limits were obtained using the assumption that the doubly charged
scalar decays exclusively into two same-sign $W$-bosons\cite{H1}\cite{H2}\cite{Chiang2}. This restricted the analyses to a small region of the parameter space. We, on the other hand, included possibilities where the doubly charged
state can also decay into a singly charged state and a $W$, as also into two singly charged states.
We find that some such cases allow higher triplet contributions to the weak boson masses
than come with the same-sign $W$-pair decay alone.

\item We have included the possibility of $v_\chi \ne v_\xi$ , i.e. of broken custodial SU(2).
The consequent shift in parameter values in the scalar potential is not fully calculable
unless one knows the UV completion of the theory, which has to be at relatively low scales
if the triplet-dominated states have to make any difference to phenomenology\cite{wud1}\cite{wud2}\cite{Chiang1}. Using
some phenomenological limits on the parameter shift, we have computed the changes in contributions
to the rates of various collider phenomena, and obtained modified limits on the
`effective' triplet vev for various masses of the doubly charged states, again allowing for 
single-channel as well as two-and three-channel decays of the latter.

\item One characteristic feature of such scenarios is the existence of non-vanishing
trilinear interactions involving a charged scalar, a $W$ and a $Z$, something that is not
permissible with scalar doublets alone\cite{HWZ}. Such interactions, essentially related to the triplet
vev value, bring in additional collider phenomenology\cite{HHG}{\cite{BMT1}-\cite{hwz2}}. We have indicated the upper
limits on the strength of such interactions, for both $v_\chi = v_\xi$ and $v_\chi \ne v_\xi$.

\item In addition to various experimental limits, direct as well as indirect,
we have also included the theoretical limits (arising mostly from unitarity and occasionally
from vacuum stability) on the triplet vev, for both $v_\chi = v_\xi$ and $v_\chi \ne v_\xi$.
The limits are found to be stronger for high values of the doubly charged scalar mass. 

\item We have calculated the coupling strength of a neutral scalar that is singlet under custodial SU(2) in this
scenario to WW/ZZ\cite{hvv1}\cite{Chiang3} and $f\bar{f}$ , using as reference the corresponding interaction strengths of the SM-like scalar. The $\gamma \gamma$ and $Z\gamma$ \cite{SCALARDECAY}\cite{Chiang4} branching ratios predicted for the same scalar, have also been calculated and scanned over the parameter space. This can serve as indicators of the phenomenological potential of this neutral spin-0 state, for instance, the high-luminosity LHC.
 
\end{enumerate}

We present a brief outline of the GM scenario in section 2.  Sections 3 and 4 are devoted to the
experimental and theoretical limits and some related features of scenarios with $v_\chi = v_\xi$ and $v_\chi \ne v_\xi$, respectively. We summarise and conclude in
section 5.
\section{Brief summary of the scenario}
The scalar sector of the Georgi-Machacek model \cite{GM}\cite{CG} consists of a $Y=2$ complex triplet $\chi = (\chi^{++}, \chi^+, \chi^0)$ and a $Y=0$ real triplet $\xi = (\xi^+, \xi^0, \xi^-)$ alongwith the usual Standard model doublet. The most general potential preserving a  global $SU(2)_L\times SU(2)_R$  is given by\cite{H1}\cite{H2}\cite{Chiang5} 

\begin{eqnarray}
\label{1e}
					V&=&\frac{\mu_2^2}{2} Tr(\Phi^\dagger\Phi) + \frac{\mu_3^2}{2}       Tr(X^\dagger X) + \lambda_1[Tr(\Phi^\dagger\Phi)]^2 + \lambda_2Tr(\Phi^\dagger\Phi)Tr(X^\dagger X)
					+ \lambda_3Tr(X^\dagger X X^\dagger X) + \lambda_4[Tr(X^\dagger X)]^2 \nonumber \\
					 &-& \lambda_5Tr(\Phi^\dagger \tau^a \Phi \tau^b)Tr(X^\dagger t^a X t^b) - M_1Tr(\Phi^\dagger \tau^a \Phi \tau^b)(UXU^\dagger)_{ab} - M_2Tr(X^\dagger t^a X t^b)(UXU^\dagger)_{ab}
\end{eqnarray}
where, 
	\begin{equation}
		\Phi = \begin{pmatrix}
		\phi^{0 \star} & \phi^+\\
		\phi^- & \phi^0 
		\end{pmatrix} \ \ \
		X = \begin{pmatrix}
						\chi^{0 \star} & \xi^+ & \chi^{++}\\
						\chi^- & \xi^0 & \chi^+\\
						\chi^{--} & \xi^- & \chi^0\\
					\end{pmatrix} \ \ \
					U = \begin{pmatrix}
    -\frac{1}{\sqrt2} & 0 & \frac{1}{\sqrt2}\\
    -\frac{i}{\sqrt2} & 0 & -\frac{i}{\sqrt2}\\
    0 & 1 & 0\\
    \end{pmatrix}
	\end{equation}
	
$\tau^a$ s and $t^a$ s are the $2\times2$ and $3\times3$ $SU(2)$ generators respectively.\newline
With this potential the tadpole conditions always have a solution at $v_{\chi} = v_{\xi}$.
The $SU(2)_L$ and the $U(1)$ subgroup of $SU(2)_R$ are gauged and they break down to $U(1)_{EM}$ via the Higgs mechanism\cite{Kibble,HM}. After electroweak symmetry breaking the vacuum is invariant under the diagonal subgroup of $SU(2)_L\times SU(2)_R$ which is called the custodial SU(2) and ensures $\rho = 1$ at the tree level. 
At $v_\chi = v_\xi$, $v_\phi^2 + 8v_\chi^2 = v^2 = (246.22 \ GeV)^2$. The tadpole conditions that determine the doublet and the triplet vev in terms of parameters in the potential, are given by\cite{H1}\cite{H2}
\begin{eqnarray}
\label{min1}
\mu_2^2 + 4\lambda_1 v_\phi^2 + 3 (2\lambda_2-\lambda_5 ) v_\chi^2 - \frac{3}{2}M_1 v_\chi = 0 \\
\label{min2}
3\mu^2_3 v_\chi + 3 (2\lambda_2 - \lambda_5 ) v_\phi^2 v_\chi + 12 (\lambda_3 + 3\lambda_4 ) v_\chi^3 - \frac{3}{4}M_1 v_\phi^2 - 18M_2 v_\chi^2 = 0
\end{eqnarray}

Since the potential retains a custodial SU(2) after EWSB the physical states will belong to the irreducible representation of this group. Hence particles belonging to a particular multiplet will be degenerate in mass. Here one has a 5-plet, two 3-plet and two singlet states. One of the 3-plets contains the Goldstone modes that become the longitudinal components of the weak gauge bosons. The 5-plet states are given by,
\begin{center}
    $H_5^{++} = \chi^{++}$, \ \ $H_5^+ = \frac{\chi^+ - \xi^+}{\sqrt{2}}$
\end{center}    
and a CP even neutral scalar,
\begin{center}
    $H_5^0 = \sqrt{\frac{2}{3}}\xi^0 - \sqrt{\frac{1}{3}}\chi^{0,r}$
\end{center}
The physical 3-plet consists of
\begin{center}
    $H_3^{+} = -s_H\phi^+ + c_H\frac{\chi^+ +  \xi^+}{\sqrt{2}}$
\end{center}    
    and a CP odd neutral scalar,
\begin{center}    
    $H_3^0 = -s_H\phi^{0,i} + c_H\chi^{0,i}$
\end{center}
where,
\begin{center}
  $s_H = \frac{2\sqrt{2}v_\chi}{v}$ and $c_H = \frac{v_\phi}{v}$.    
\end{center}
The masses of these 5-plet and 3-plet scalars are,
\begin{eqnarray}
\label{5m}
{m^2_{H_5^{\pm\pm}}} = {m^2_{H_5^{\pm}}} = {m^2_{H_5^0}} = m_5^2 &=& \frac{M_1}{4v_\chi}v_\phi^2 + 12M_2v_\chi + \frac{3}{2}\lambda_5v_\phi^2 + 8\lambda_3v_\chi^2 \\
\label{3m}
{m^2_{H_3^\pm}} = {m^2_{H_3^0}} = m_3^2 &=& (\frac{M_1}{4v_\chi} + \frac{\lambda_5}{2})v^2
\end{eqnarray}
The singlet states are given by,
\begin{center}
    $H^0 = \phi^{0,r}$ , $H^{0\prime} = \sqrt{\frac{1}{3}}\xi^0 + \sqrt{\frac{2}{3}}\chi^{0,r} $
\end{center}
The physical states are combinations of these two and given by,
\begin{center}
    $h = \cos{\alpha}\phi^{0,r} - \sin{\alpha}H^{0\prime}$ , $H = \sin{\alpha}\phi^{0,r} + \cos{\alpha}H^{0\prime}$
\end{center}
The angle $\alpha$ depends on the $2 \times 2$ CP-even custodial-singlet scalar mass matrix. The elements of the mass matrix are,
\begin{eqnarray}
\mathcal{M}_{11}^2 &=& 8\lambda_1 v_\phi^2 \\
\mathcal{M}_{12}^2 &=& \frac{\sqrt{3}}{2} v_\phi[-M_1 + 4(2\lambda_2 - \lambda_5)v_\chi] \\
\mathcal{M}_{22}^2 &=& \frac{M_1 v_\phi^2}{4 v_\chi} - 6M_2v_\chi + 8(\lambda_3 + 3\lambda_4)v_\chi^2 \\
\tan2\alpha &=& \frac{2\mathcal{M}_{12}^2}{\mathcal{M}_{22}^2 - \mathcal{M}_{11}^2}
\end{eqnarray}
Here we have set $h$ to be the 125 GeV scalar and denoted the mass of H as $m_H$ which can be larger as well as smaller than 125 GeV depending on the parameters of the scalar potential. Since the most stringent constraint on the parameter space of this model comes from the collider searches of the doubly charged Higgs, its branching ratios in different channels play a crucial role. Hence we will treat the mass of the 5-plet state, 3-plet state and the 125 GeV scalar as our input parameter and will trade off three potential parameters in terms of them. Thus the final set of input parameter for our study is ${m_5, m_3,s_H,\lambda_2,\lambda_3,\lambda_4,M_2}$.\newline
In this model the 5-plet states couple to vector boson pairs as opposed to the 3-plet states which only couple to fermions.
Let us in this context define the quantites, $\zeta_W$, $\zeta_Z$  and $\zeta_f$,   which reflect the strengths of the $W,Z$  and
fermionic coupling,  of H normalized to the similar couplings of the SM Higgs respectively\cite{HWZ}.
\begin{eqnarray}
\label{zz}
\zeta_Z &=& \frac{g^{GM}_{HZZ}}{g^{SM}_{hZZ}} = \frac{\sqrt{2}}{v}{\Sigma_k \beta_{ki}v_k^cY_k^2} \\
\label{zw}
\zeta_W &=& \frac{g^{GM}_{HWW}}{g^{SM}_{hWW}} = {\frac{\sqrt{2}}{v}\Sigma_k \beta_{ki}(v_k^cY_k + v_k^r \frac{1}{4}\sqrt{n_k^{r^{2}} - 1} } ) \\
\label{zf}
\zeta_f &=& \frac{g^{GM}_{Hf\bar{f}}}{g^{SM}_{hf\bar{f}}} = \beta_{12} {v}/{v_\phi}
\end{eqnarray}
\noindent where $\beta$ is the matrix that rotates $(h,H,H_5^0)$ to $(\phi^r,\chi^r,\xi^r)$ and summation is over all the scalar multiplets participating in EWSB. In eq. \eqref{zz} and \eqref{zw} $i = 2$. $n_k^r = 3$ for the single real triplet $\xi$, while $v_k^c$ and $v_k^r$ are the vevs of the complex and real multipltes respectively.
As long as the custodial symmetry is preserved at both the potential and the vacuum level, $\zeta_W = \zeta_Z = \zeta_V = \sin\alpha\frac{v_\phi}{v} + \frac{8}{\sqrt{3}}\cos\alpha\frac{v_\chi}{v}$ and breaking of custodial symmetry will introduce a splitting between them. Also with custodial symmetry $\beta_{12} = \sin{\alpha}$ and $\beta_{13} = 0$. Another remarkable feature of this model is the presence of the non-zero $H^{\pm}W^{\mp}Z$ coupling. With custodial symmetry intact only $H_5^{\pm}$ has this kind of coupling and when the symmetry is broken a non-zero $H_3^{\pm}W^{\mp}Z$ vertex strength starts to appear. For the custodial symmetric case, this coupling strength is a simple scaling to $s_H$ by a factor of $\frac{2 M_W^2}{V c_W}$, where $c_W$ is the cosine of the Weinberg angle. But when the custodial symmetry is broken this coupling takes a complicated form. For this case let us define two quantities,
\begin{eqnarray}
\kappa_{H_5^{\pm}W^{\mp}Z} &=& \frac{g^{NC}_{H_5^{\pm}W^{\mp}Z}}{[g^{CS}_{H_5^{\pm}W^{\mp}Z}]_{s_H = 1}} \\
\kappa_{H_3^{\pm}W^{\mp}Z} &=& \frac{g^{NC}_{H_3^{\pm}W^{\mp}Z}}{[g^{CS}_{H_5^{\pm}W^{\mp}Z}]_{s_H = 1}}
\end{eqnarray}
where, $g^{NC}_{H_5^{\pm}W^{\mp}Z}$ and $g^{NC}_{H_3^{\pm}W^{\mp}Z}$ denotes the ${H_5^{\pm}W^{\mp}Z}$ and ${H_3^{\pm}W^{\mp}Z}$ interaction vertex strengths for $v_\chi \ne v_\xi$ case and $g^{CS}_{H_5^{\pm}W^{\mp}Z}$ denotes the ${H_5^{\pm}W^{\mp}Z}$ interaction vertex strength for custodial symmetric case.
The expressions of $g_{H_5^{\pm}W^{\mp}Z}$, following a general derivation given \cite{HWZ}, are
\begin{equation}
\label{hwz}
    g_{H_i^{\pm}W^{\mp}Z} = \frac{g^2}{\sqrt{2}\cos{\theta_W}}\Sigma_k \alpha_{ki}(f_k^cv_k^c + f_k^rv_k^r)
\end{equation}
where $\alpha$ is the $3 \times 3$ matrix that diagonalises  the singly charged scalar mass matrix in $(\phi^+, \chi^+, \xi^+)$ basis, $f_k^c = \sqrt{n_k^c - 1}(\cos^2{\theta_W} - Y_k)$, $f_k^r = \frac{1}{2}\sqrt{{n_k^r}^2 - 1}\cos^2{\theta_W}$. Here $f_k^c$ corresponds to $k = 1,2$ comprising $\phi$ and $\chi$, with $n_k^c = 2$ and 3 respectively. Similarly $f_k^r$ corresponds to $\xi$ with $k=3$ and $n_k^r = 3$. In the situation where $v_\chi =  v_\xi$ the $H^{\pm}W^{\mp}Z$  coupling is possible for $H_5^+$ only and then the expression in eq. \eqref{hwz} reduces to,
\begin{equation}
\label{hwzcs}
    g_{H_5^+W^-Z} = \frac{-gm_W s_H}{\cos{\theta_W}}
\end{equation}

It should be noted that the assertion $v_\chi = v_\xi$, based on eq. \eqref{1e} and the invariance of the potential under the custodial $SU(2)$, can be exactly ensured at the tree level only. However, the kinetic energy terms explicitly breaks the custodial $SU(2)$ via $U(1)_Y$ interactions. This at higher orders violates the custodial symmetry of the potential, resulting $v_\chi \ne v_\xi$ (Unless there is fine-tuning). It is therefore useful to also estimate the modifications to the constraints on the parameter space, when the real and complex triplet vevs are unequal. In particular, the various points itemized in the previous section needs to be revisited for such a situation. The higher order modifications , however, leads to large contributions to mass parameters in the potential, which in general shift them to the upper limit of validity of the GM scenario. Therefore in case the triplet sector is expected to have bearing on accessible phenomenology, the GM model will have to pass the baton to some over-seeing scenario not too far above the TeV scale, which controls the divergent corrections\cite{susygm1} - \cite{susygm3}. This essentially brings in new interactions, symmetries or degrees of freedom. Thus it is difficult to compute the modified potential in terms of the GM parameters alone. Keeping this in mind, we have studied the constraints for $v_\chi \ne v_\xi$ in section 4, using some illustrative scenarios introduced in a phenomenological manner.

\section{Constraints for $v_\chi = v_\xi$}
We start by updating and extending the existing constraints on the parameter space of the GM scenario. Before stating explicitly
where we have gone beyond the studies already performed in this direction \cite{H1}\cite{H2}, a few general comments are in order.

The limits obtained in most works including ours are in terms of the mass
of the 5-plet of  the custodial $SU(2)$\cite{H1}\cite{H2}\cite{Chiang8}. In particular, it is meaningful to find the allowed upper limits of the quantity $s_H$,
which is a measure of
the triplet vev relative to the `effective' vev $v$ driving EWSB, for various values of the doubly charged scalar mass. Stronger
bounds of this kind come from direct searches in the 5-plet sector rather than those in the 3-plet  sector, because
the production rates for this sector at the LHC are suppressed by $\frac{8v_\chi^2}{v_\phi^2}$ relative to the SM Higgs. Now,
  members of the 3-plet decay in fermionic channels only, where decays into tau(s) is one's most favourable option in terms
  of cleanliness of the signal. With the production rate already suppressed, the final states containing taus offer
  rather poor statistics, so that the limits cannot be significant.

   As far as the 5-plet sector is concerned, the neutral or the singly charged member yields weaker limits  on $s_H$ than what one obtains for  the doubly charged  component. For $H^{++}_5$,
  on the
  other hand there are in principle various decay channels like $W^{\pm} W^{\pm}, H^{\pm}_3 W^{\pm}, H^{\pm}_3 H^{\pm}_3$ as also
  $\ell^{\pm} \ell^{\pm}$ when $\Delta L = 2$ Yukawa couplings exist. Same-sign dileptons, wherever available out of such final
  states,
  make up for the generally existing $s_H$-suppression by virtue of the relatively `clean' visible peaks.
  That is why it is more profitable to attempt setting  limits on $s_H$ against $m_{H_5^{\pm\pm}}$, keeping of course in view
  other considerations going beyond direct searches.

  Among the channels listed above, $H^{++}_5 \rightarrow \ell^+ \ell^+$ is undoubtedly the most spectacular signal, where one
  notices a peak in the invariant mass distribution of the same-sign dileptons\cite{GMdilep}. However, given the potential contribution
  of this scenario to neutrino masses via the Type-2 seesaw mechanism, a significant branching ratio for
  $H^{++}_5 \rightarrow \ell^+ \ell^+$ will require $v_\chi \le 10^{-10}$ GeV
  \cite{trip1}\cite{TRIP3}. This however is a situation far removed from those
  which are our main focus here, namely, the viability of substantial role of triplets in EWSB. We therefore keep the $\ell^{\pm} \ell^{\pm}$
  decay channel out of our consideration.

  Among the three remaining channels mentioned above,   $H^{\pm}_3 W^{\pm}$ and  $H^{\pm}_3 H^{\pm}_3$ will have the 3-plet
  charged scalars mostly decaying into quarks , in which case the leptons, even if produced, are likely to be degraded
  in cascades. Their detection is therefore relatively inefficient, and no analysis exists on these two channels so far.
  They, however, may serve to suppress the branching ratio into $W^{\pm} W^{\pm}$, thereby weakening the limits and allowing
  higher values of $s_H$ for a given $m_5$, at least from direct search at the LHC. We therefore extend the existing
  analyses for not only the single decay channel $H^{++}_5 \rightarrow W^{\pm} W^{\pm}$ but also with either one or two of the
  additional channels mentioned above. In addition, the following constraints are taken into account:

  \begin{itemize}
  \item Indirect constraints\cite{hindr}, primarily from the experimental bounds  on the rates for
    $b \rightarrow s\gamma$\cite{Bsgam} and $B_s \rightarrow \mu^+ \mu^-$\cite{Bmumu},
    and also from oblique electroweak parameters (of which the limit from $T$, or equivalently the $\rho$-parameter,
    is explicitly used at every step).
  \item Theoretical constraints such as perturbative unitarity, vacuum stability, and the requirement that
  the custodial symmetry preserving vacuum corresponds to the global minimum of the potential\cite{UNI}\cite{DECOUP}.

\item Whatever constraints come withinthe scope of the package GMCALC\cite{gmcalc} have been made use of. We have updated them, using our own code for VBF with 137~$fb^{-1}$ and DY with 139~$fb^{-1}$. The consistency of the code developed by us has been checked against GMCALC in the appropriate limits.  

\item All otherconstraints from the searches for additional neutral searches obtained using the code HiggsBounds\cite{HB5}.

\item  The requirement that the signal strength of the 125-GeV scalar in all channels is consistent with LHC data.
      For this, the code HiggsSignals\cite{HS2} has been made use of. 
  \end{itemize}

\begin{figure}[!htb]
     \begin{subfigure}[b]{0.25\textwidth}
         \includegraphics[width=1.3\textwidth]{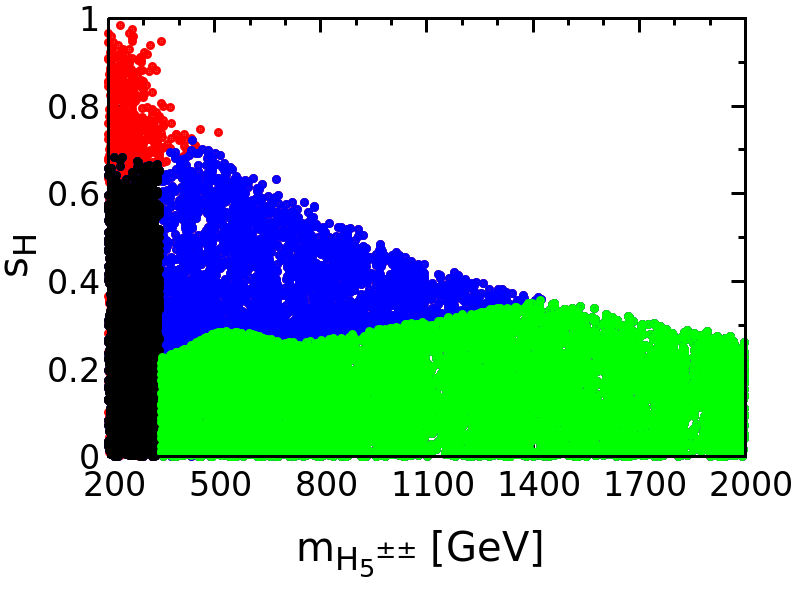}
         \caption{}
         \label{1(a)}
     \end{subfigure}
     \hspace{0.06\textwidth}
     \begin{subfigure}[b]{0.25\textwidth}
         \includegraphics[width=1.3\textwidth]{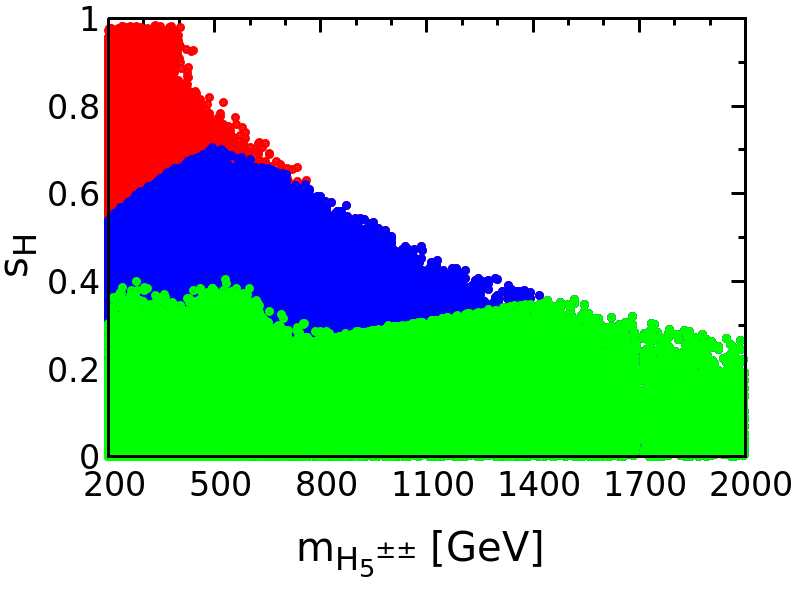}
         \caption{}
         \label{1(b)}
     \end{subfigure}
     \hspace{0.06\textwidth}
     \begin{subfigure}[b]{0.25\textwidth}
         \includegraphics[width=1.3\textwidth]{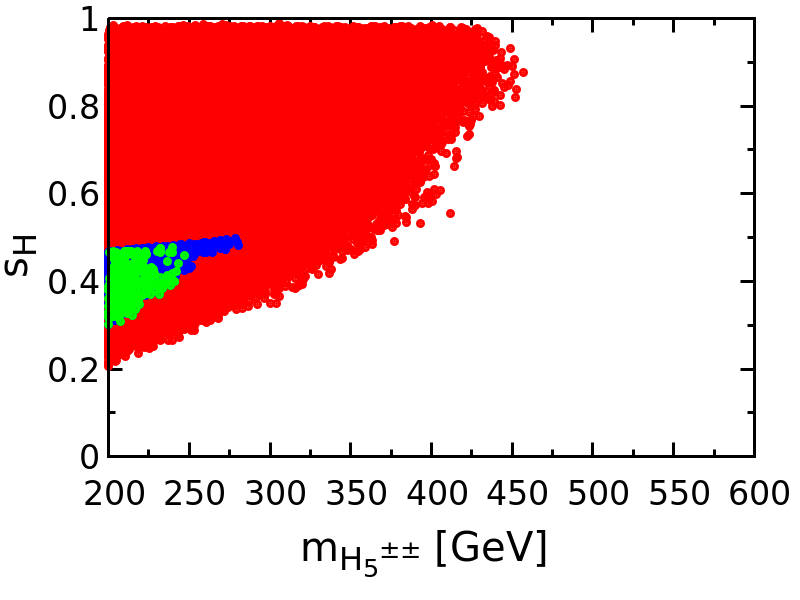}
         \caption{}
         \label{1c}
     \end{subfigure}
    \caption{ Constraints on the $m_{H_5^{\pm\pm}} - s_H$ plane. 1(a),(b) and (c) correspond respectively to the situations of single, double and triple channel decay of $H_5^{++}$. Red, black and blue regions are excluded by indirect constraints, DY search of $H_5^{++}$ and VBF of $H_5^{++}$ respectively. Green regions are allowed by all constraints.}
    \label{1}
\end{figure}

\begin{figure}[!htb]
     \begin{subfigure}[b]{0.25\textwidth}
         {\centering
         \includegraphics[width=1.3\textwidth]{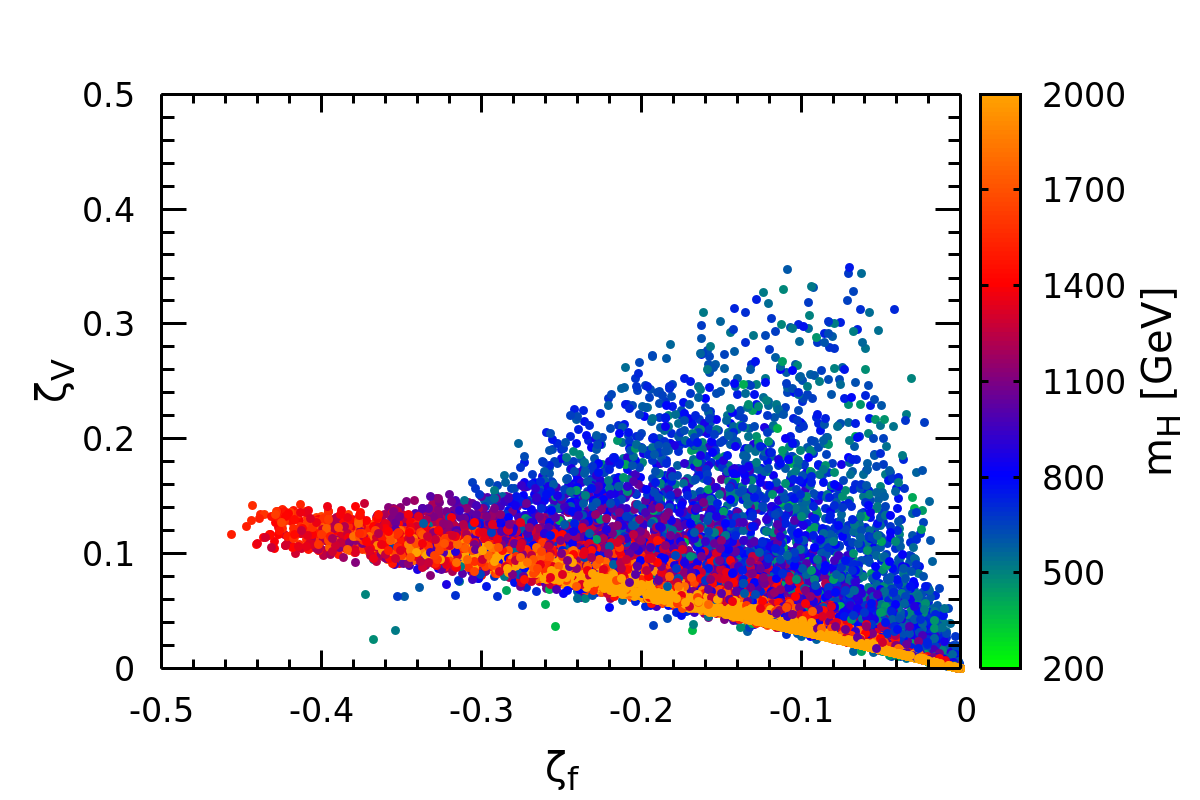}
         \caption{}
         \label{2a}}
     \end{subfigure}
     \hspace{0.06\textwidth}
     \begin{subfigure}[b]{0.25\textwidth}
         {\centering
         \includegraphics[width=1.3\textwidth]{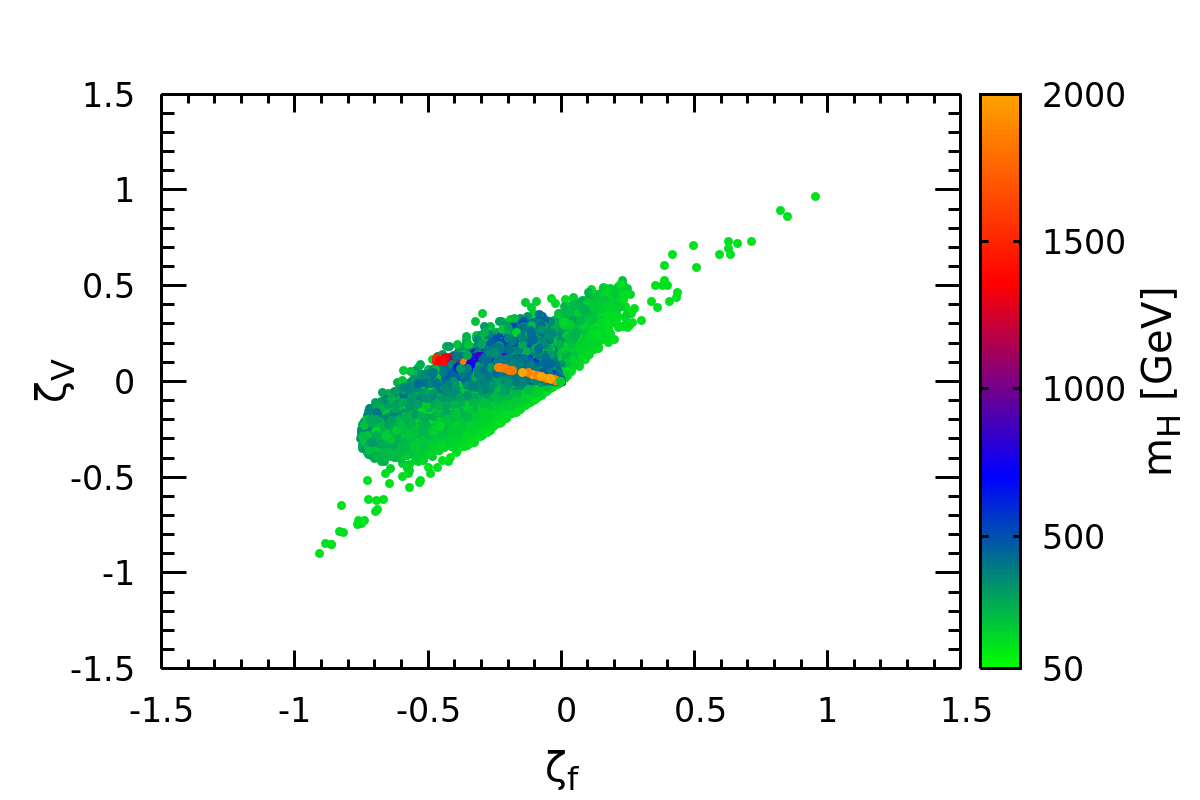}
         \caption{}
         \label{2b}}
     \end{subfigure}
     \hspace{0.06\textwidth}
     \begin{subfigure}[b]{0.25\textwidth}
         {\centering
         \includegraphics[width=1.3\textwidth]{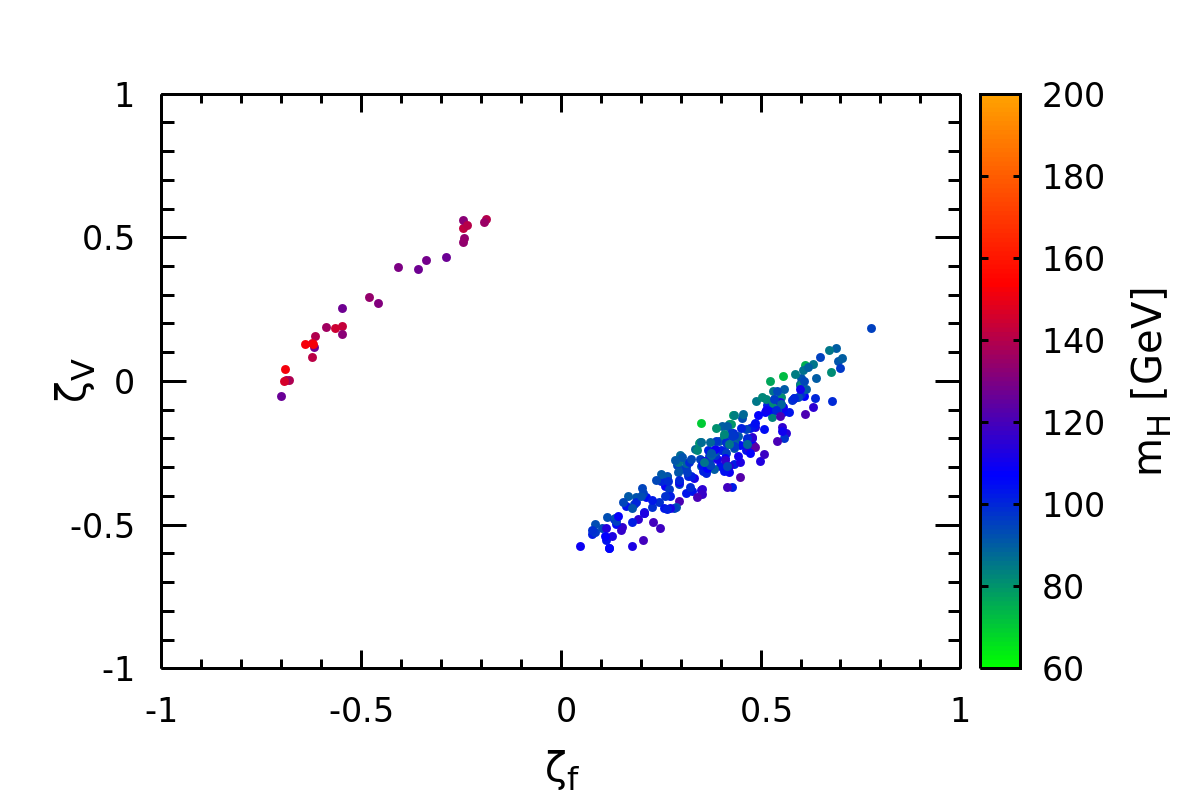}
         \caption{}
         \label{2c}}
     \end{subfigure}
     
     \centering
     \begin{subfigure}[b]{0.25\textwidth}
         {\centering
         \includegraphics[width=1.3\textwidth]{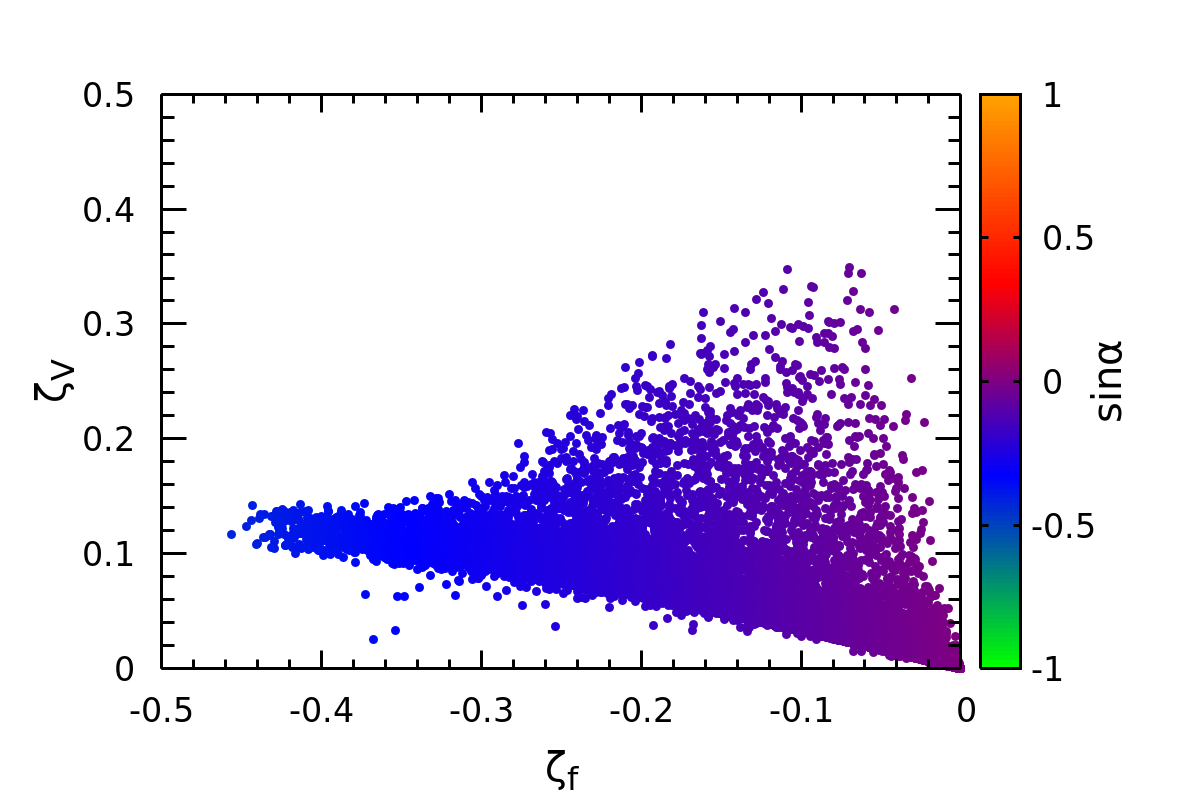}
         \caption{}
         \label{2d}}
     \end{subfigure}
     \hspace{0.06\textwidth}
     \begin{subfigure}[b]{0.25\textwidth}
         {\centering
         \includegraphics[width=1.3\textwidth]{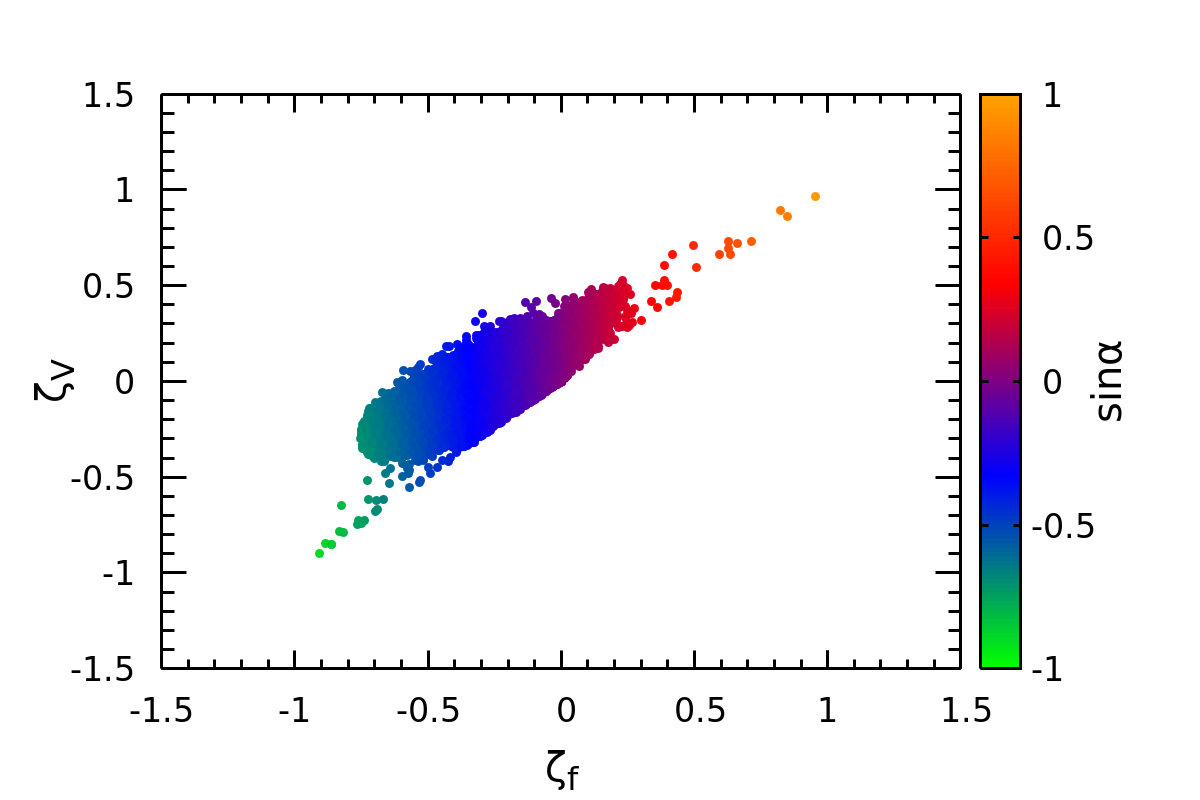}
         \caption{}
         \label{2e}}
     \end{subfigure}
     \hspace{0.06\textwidth}
     \begin{subfigure}[b]{0.25\textwidth}
         {\centering
         \includegraphics[width=1.3\textwidth]{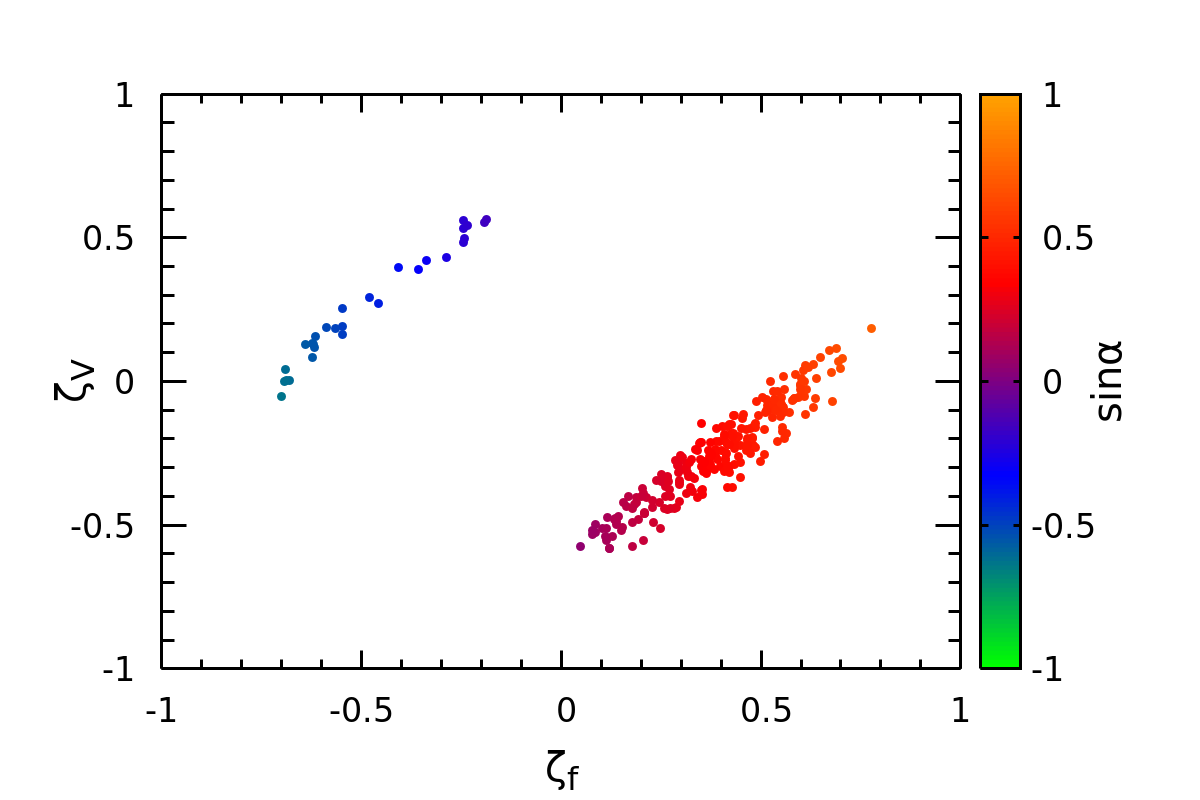}
         \caption{}
         \label{2f}}
     \end{subfigure}

     \caption{Correlated modification factors in coupling strengths for the neutral scalar H to weak boson pairs and fermion pairs with respect to the corresponding couplings of the SM Higgs for single channel(a and d), double channel(b and e) and triple channel(c and f) decays of $H_5^{++}$. The colour axis in (a),(b) and (c) correspond to $m_H$, the mass of H while in (d),(e) and (f) it corresponds to $\sin{\alpha}$, the $SU(2)_L$ doublet content of H.}
     \label{2}
\end{figure}

\begin{figure}[!htb]
\label{3}
     \centering
     \begin{subfigure}[b]{0.25\textwidth}
         \centering
         \includegraphics[width=1.3\textwidth]{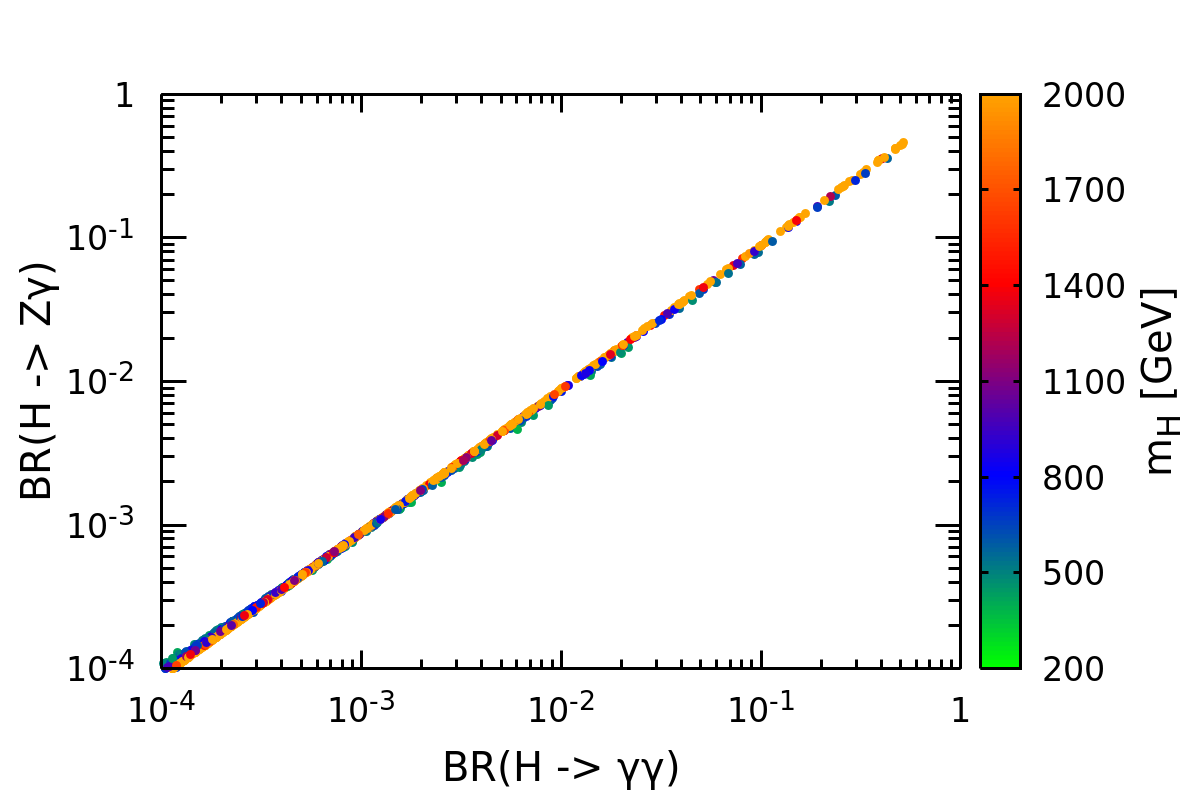}
         \caption{}
         \label{3a}
     \end{subfigure}
     \hspace{0.06\textwidth}
     \begin{subfigure}[b]{0.25\textwidth}
         \centering
         \includegraphics[width=1.3\textwidth,height=0.85\textwidth]{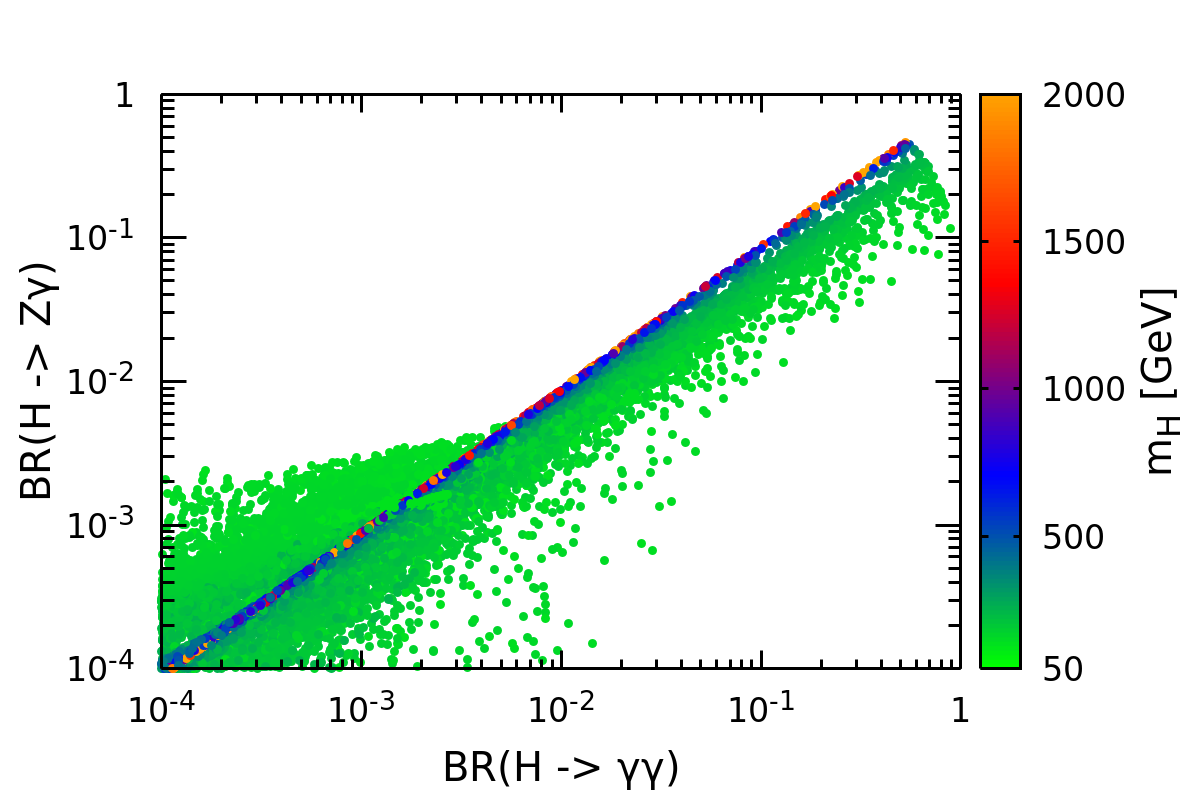}
         \caption{}
         \label{3b}
     \end{subfigure}
     \hspace{0.06\textwidth}
     \begin{subfigure}[b]{0.25\textwidth}
         \centering
         \includegraphics[width=1.3\textwidth]{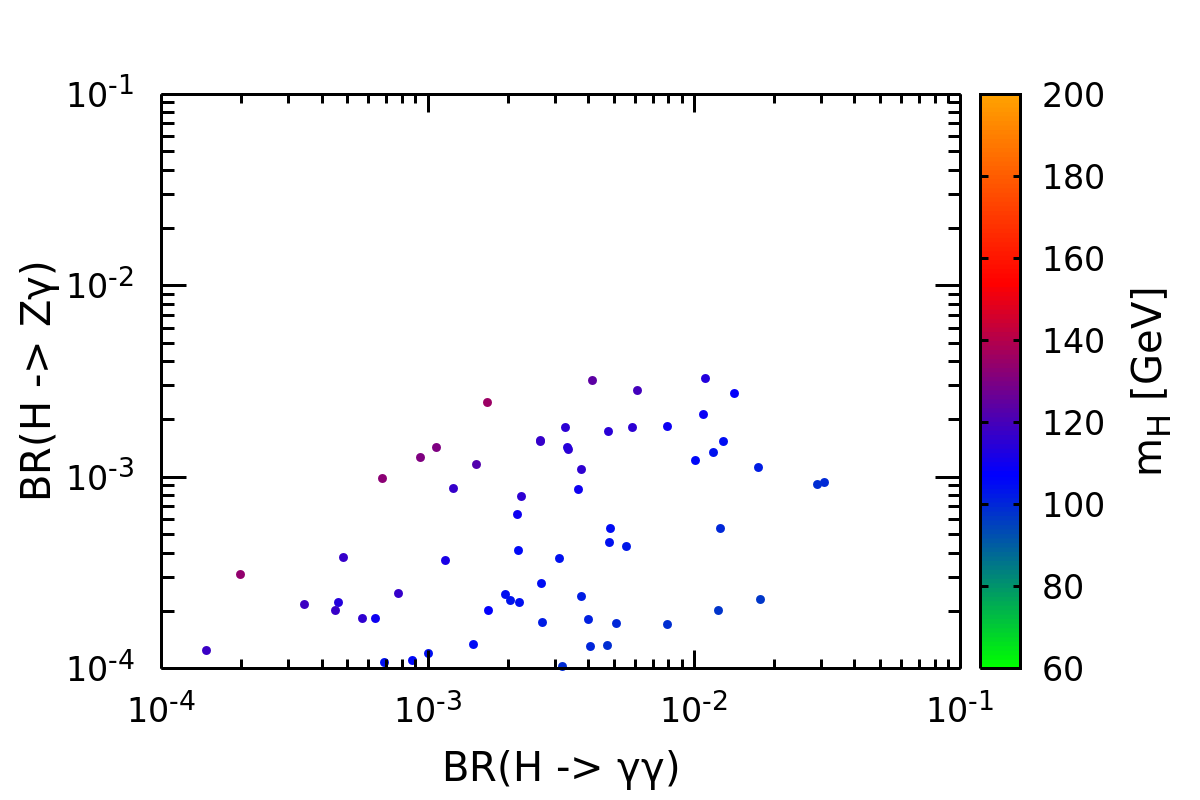}
         \caption{}
         \label{3c}
     \end{subfigure}

     \centering
     \begin{subfigure}[b]{0.25\textwidth}
         \centering
         \includegraphics[width=1.3\textwidth]{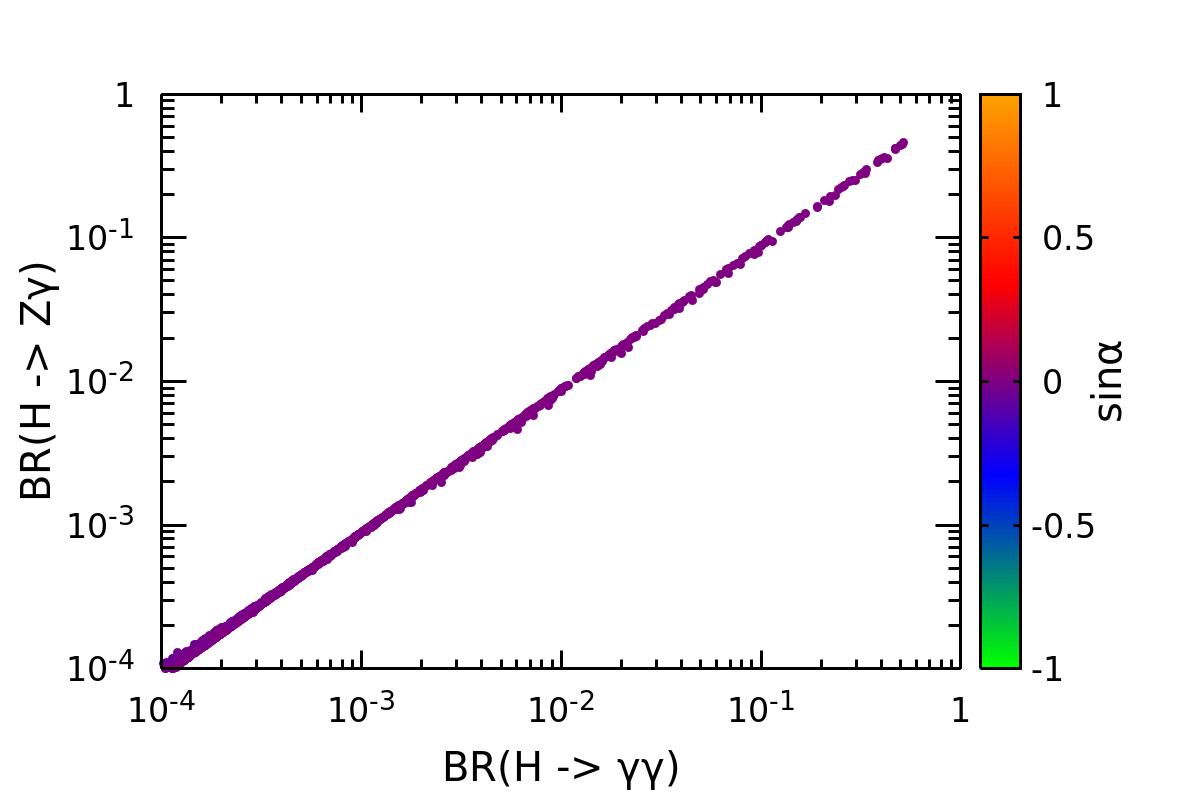}
         \caption{}
         \label{3d}
     \end{subfigure}
     \hspace{0.06\textwidth}
     \begin{subfigure}[b]{0.25\textwidth}
         \centering
         \includegraphics[width=1.3\textwidth,height=0.85\textwidth]{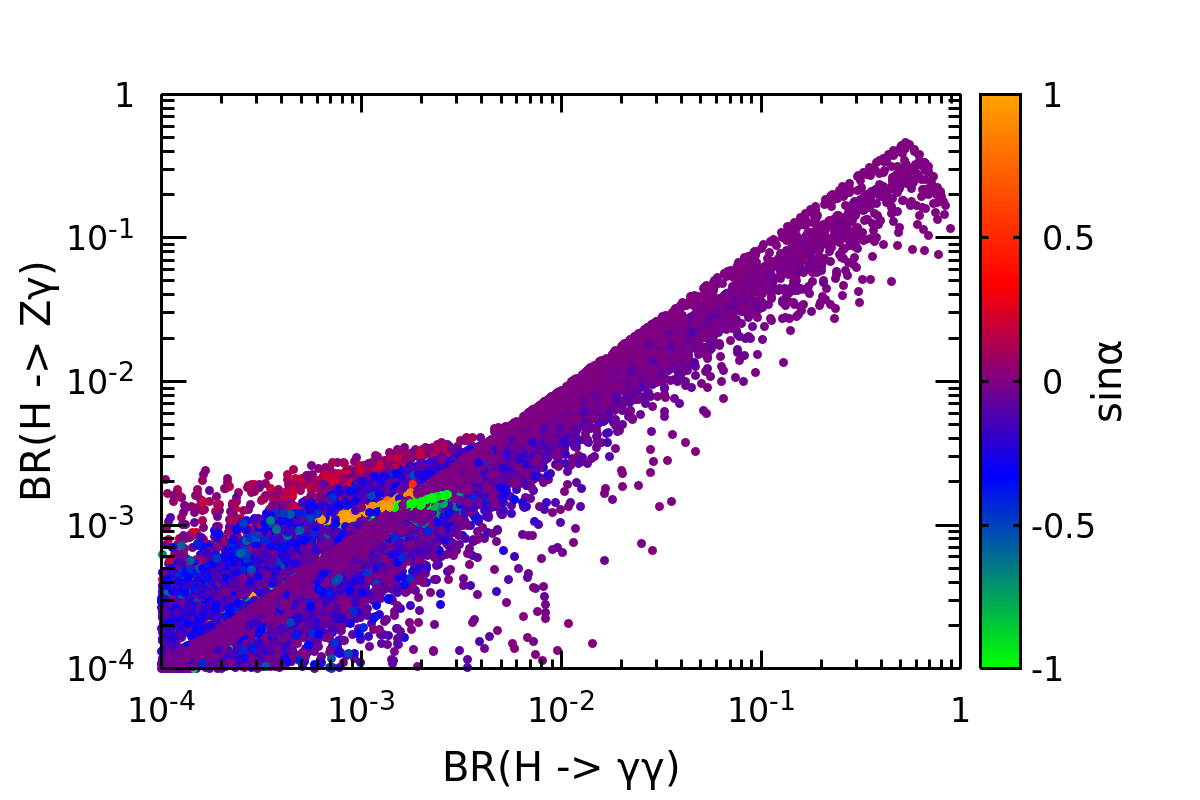}
         \caption{}
         \label{3e}
     \end{subfigure}
     \hspace{0.06\textwidth}
     \begin{subfigure}[b]{0.25\textwidth}
         \centering
         \includegraphics[width=1.3\textwidth]{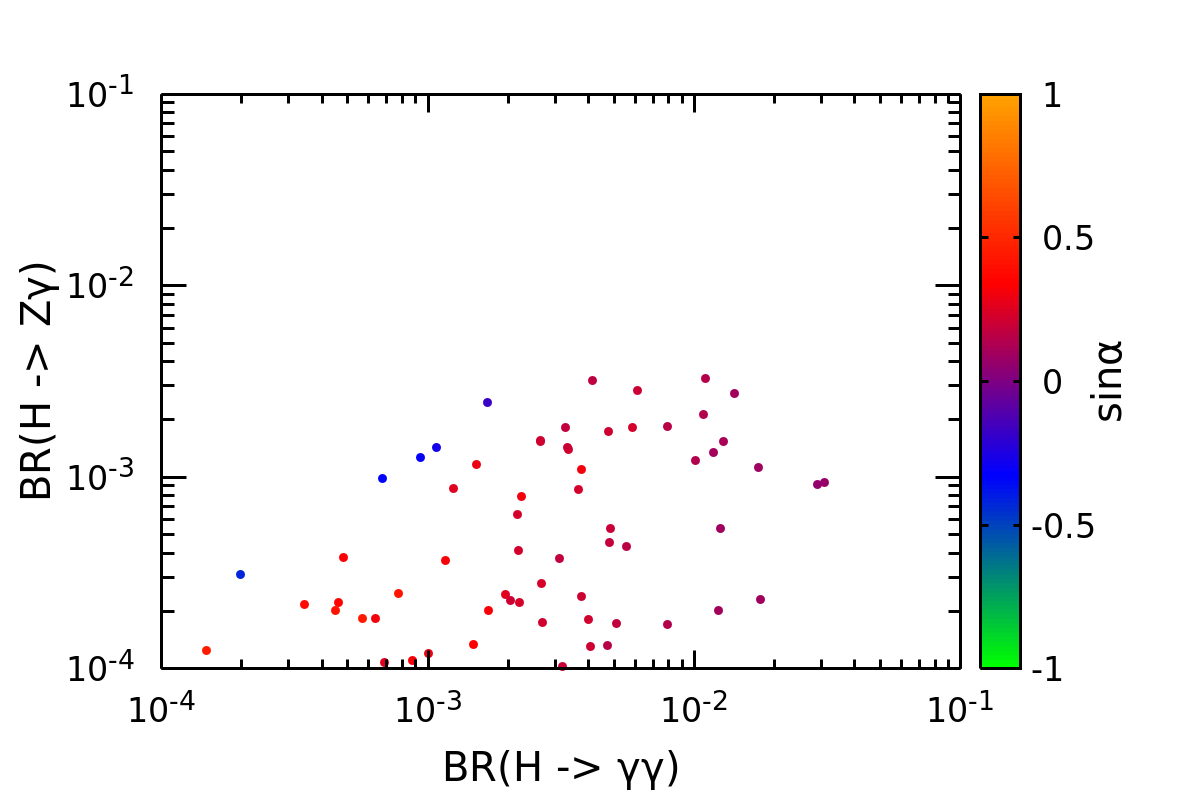}
         \caption{}
         \label{3f}
     \end{subfigure}
     \caption{Correlated branching ratios for $H \rightarrow Z \gamma$ and $H \rightarrow \gamma \gamma$ for single(a and d), double(b and e) and triple-channel(c and f) decay of $H_5^{++}$. The colour axis in (a),(b) and (c) correspond to $m_H$, the mass of H while in (d),(e) and (f) it correspond to $\sin{\alpha}$, the $SU(2)_L$ doublet content of H.} 
     \label{3}
\end{figure}

In obtaining the scatter plots presented in Figures \ref{1}-\ref{3}, the quantities whose specific values have entered 
as independent variables are
$m_5, m_3$ and $s_H$, all other parameters in the GM scalar potential being subjected to
an {\it unbiased scan}, subject to the requirements of unitarity, vacuum stability and the demand that
the EWSB vacuum corresponds to global minimum of the potential.  The mass parameter ($m_5$) corresponding to the
custodial 5-plet has been varied in the range $200$ GeV - $2$ TeV, subject to the constraints mentioned above.
The 3-plet mass $m_3$ is also subjected similar constraints, with  its maximum allowed values being set according
the different kinematic situations described in the next paragraph. In cases where the 3-plet is never produced
in decays of the doubly charged scalar $H^{++}_5$, the scan  has been extended upto $m_3 = 5$ TeV. 
The quantity $s_H$ has been scanned over the entire allowed  range of positive values, namely, $[0,1]$.  

Fig. \ref{1} contains the updated constraints obtained through our analysis, while the coupling strength modifications
over the  allowed region are shown in Fig. \ref{2}, taking as illustration the custodial-singlet
neutral state $H$. Figures \ref{1}(a), (b), (c) correspond, respectively to single channel($H_5^{++} \rightarrow W^+ W^+$), double channel($H_5^{++} \rightarrow W^+ W^+,  H_3^+ W^+$) and triple channel($H_5^{++} \rightarrow W^+ W^+,  H_3^+ W^+, H_3^+ H_3^+$) decays allowed for the doubly charged state.
In Fig. \ref{2}, the left, central and right coloumns fall in these three kinematic categories.

Fig. \ref{3} illustrate the modified branching ratios of the CP-even, custodial-singlet scalar $H$ in the
$Z\gamma$ and $\gamma\gamma$ channels. This may be compared with the corresponding branching ratios
of the SM Higgs boson.

While the three sets of figures and and their corresponding captions are self-explanatory, the following features
exhibited by them need to commented upon:

\begin{itemize}
\item The additional integrated luminosity accumulated upto 137 $fb^ {-1}$ puts additional constraints
  on $s_H$ via the VBF channel. Thus the  upper limit in the case of single-channel $H_5^{++}$  decay
  becomes more stringent compared to that found in \cite{H2}, especially for masses exceeding 500 GeV.

\item When the decay channel into   $H_3^+ W^+$ opens up, the limits from VBF are relatively relaxed, and can go up to
  $s_H \approx 0.4$ at 500 GeV, where the earlier limit is 0.28 \cite{H1}\cite{H2}.

\item All the three aforementioned decays open up for a rather limited region of the parameter space, which ends
  around $m_5 \approx 300 GeV$. This is because it is not possible beyond this mass to satisfy equations
  5 and 6 simultaneously, without creating an unacceptable tension between the potential minimisation conditions
  and unitarity limits on the quartic couplings. However, as expected, higher values of $s_H$ (upto about 0.44) get
  allowed within this range, as compared to both the single-and double-channel decay situations. Looking at it
  from another angle, the higher $m_5$ is, the more difficult it is to keep $m_3$ small enough for the above channel
  to open up.

\item The allowed values of $s_H$ goes down significantly for large $m_5$. The reason for this is
  understood if one looks at equation \eqref{min1} and \eqref{5m}. A high mass for $H_5^{++}$ is largely governed by $M_1$. Unless
  the triplet vev $v_\chi$ (and therefore $s_H$) is small, large $M_1$ would enhance some of the quartic couplings,
  ending up in unitarity violation.

\item The Drell-Yan channel has also been taken into account for 139~fb$^{-1}$ of  integrated luminosity. For the single
  channel case, this practically excludes the mass range 200 - 300 GeV for the doubly charged scalar \footnote{For masses below 200 GeV, the DY limits\cite{DYREF1}\cite{DYREF2} are weaker compared to those from VBF}
  The DY limits however, are progressively weaker for the two-and three-channel cases, as the sensitivity goes down proportionally
  with the squared branching ratio into $W^+ W^+$.

\item The code HiggsBounds has been used to constrain the particle spectrum from neutral Higgs searches at the LHC
  as well as LEP. However, the limits thus obtained are weaker than those from VBF\cite{VBF2}\cite{VBF1}. As for the constraints from HiggsSignals, the allowed regions shown the Fig. \ref{1(a)} and \ref{1(b)}  are consistent at 95\% C.L. However, the low $H_3^+$ masses relevant to the three-channel case tend to create more tension with the measurements of the diphoton signal strength for the 125-GeV scalar. This explains, for instance, the paucity of allowed points in Fig. \ref{2c} where 95\% C.L. consistency with HiggsSignals results are explicitly demanded.

\item Fig. \ref{2}  shows the modifications of the couplings for the 
  custodial-singlet neutral scalar $H$
  to fermion and gauge boson pairs, as compared to those for the SM Higgs boson. As is particularly evident in Fig. \ref{2d}-\ref{2f}, the modifications are indicative of the SU(2) doublet content of this state, and can have bearings on its detectability at colliders. It should be
  noted that this content can be distinctly higher in the double-and triple-channel cases, although the regions of
  the parameter space answering to such enhancement are rather small. 

\item As seen from Fig. \ref{3}, it is possible for both the $\gamma\gamma$ and $Z\gamma$ branching ratios 
  of
  the custodial singlet $H$ to be considerably larger than those of the SM Higgs boson. This is because of some interesting features of this scenario. First, the tree-level decays of $H$ into fermions get suppressed, since
 such decays are driven by its doublet component alone. Similarly, the role of the triplets in
 decays into gauge boson pairs is suppressed by the triplet vev. Such reduction in the share of tree-level decays results in enhancement of the
 loop-induced $\gamma\gamma$ and $Z\gamma$ branching ratios. Secondly, while the fermion loops
 involve only the doublet component, the triplet part, too, drives the gauge boson loop amplitude. 
 As a consequence, for a relatively small fraction
 of points used  in the scan, the destructive interference between the fermion and gauge boson loops
 becomes constructive. More importantly, the charged scalar loop contribution has a rather strong 
 constructive interference with the fermion loops. 
 
 As mentioned above, such an effect is noticeable for a small number of points corresponding
 to $H^{++}_5$-decays  in single-and double-channel. Enhancement is rarer in cases where 
 three decay channels open up.
  Nonetheless, this feature can have interesting
  phenomenological implications in LHC searches for this state \cite{BMT1}\cite{BMT2}\cite{BMT3}, albeit within a limited region of
  the parameter space. 
  
 \item As has been mentioned already, a striking feature of this scenario (and in general of any theory containing scalar multiplets with non-zero vev possessing different values of $Y$)  is  tree-level interaction involving a charged Higgs, a $W$-boson and a 
 $Z$-boson \cite{HHG}\cite{HWZ}, which can have interesting implications in triplet scalar phenomenology. For $v_\chi = v_\xi$,
 the strength of this interaction is proportional to $s_H$. This proportionality does not hold when the
 complex and real triplet vevs are unequal, as given in section 2. The numerical constraints on the $H^{+} W^{-}Z$ couplings in such cases are explicitly obtained in the next section.   
 \end{itemize}

\section{Constraints for $v_\chi \ne v_\xi$}

So far we have discussed the constraints corresponding to $v_\chi = v_\xi$ which is a direct
consequence of the custodial $SU(2)$ symmetry of the scalar potential . Although this symmetry
has been shown to be preserved in scalar loop corrections  \cite{CG}, it is in general
liable to be broken on inclusion of gauge couplings. This is because hypercharge interactions in
the scalar covariant derivatives break the custodial $SU(2)$.  We have studied the constraints on
model parameters, and associated issues,  in such a situation as well, as reported below.
 
The  higher-order corrections mentioned above  entail quadratically divergent terms\cite{wud2}. If the triplet scalar 
mass terms in the GM scenario have to be around the TeV-scale or less (which is the situation
where the GM model is phenomenologically significant), then some additional inputs have to come
in the form of a cut-off for the GM theory. This has prompted us to use
some phenomenological inputs for $v_\chi \ne v_\xi$. 

Our investigation pertains to cases where the theoretical limits (which in principle requires some knowledge of the overseeing
theory controlling the divergent contributions) are not substantially different from those in the corresponding
cases with $v_\xi = v_\chi$. A similar consideration applies to indirect constraints.\footnote {It should be
remembered that the inequality of the two kinds of triplet vevs results in  $H_3^+ W^- Z$ interactions, too,
which alters the limits from precision electroweak observables. Similarly, $H^\pm_5$ now develops  small
fermion couplings.}

\begin{figure}[!htb]
     \centering
     \begin{subfigure}[t']{0.25\textwidth}
         \centering
          \includegraphics[width=1.3\textwidth]{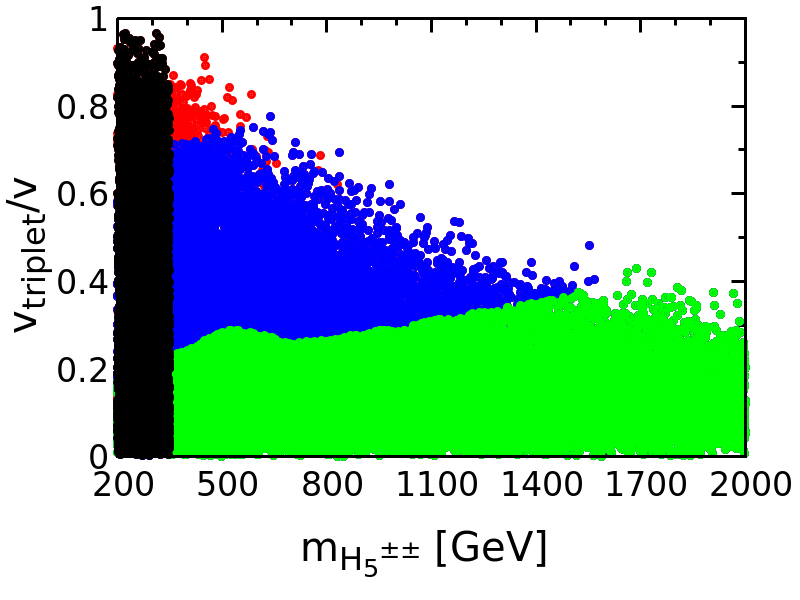}
          \caption{}
        \label{4a}
     \end{subfigure}
     \hspace{0.06\textwidth}
     \centering
     \begin{subfigure}[t']{0.25\textwidth}
         \centering
          \includegraphics[width=1.3\textwidth]{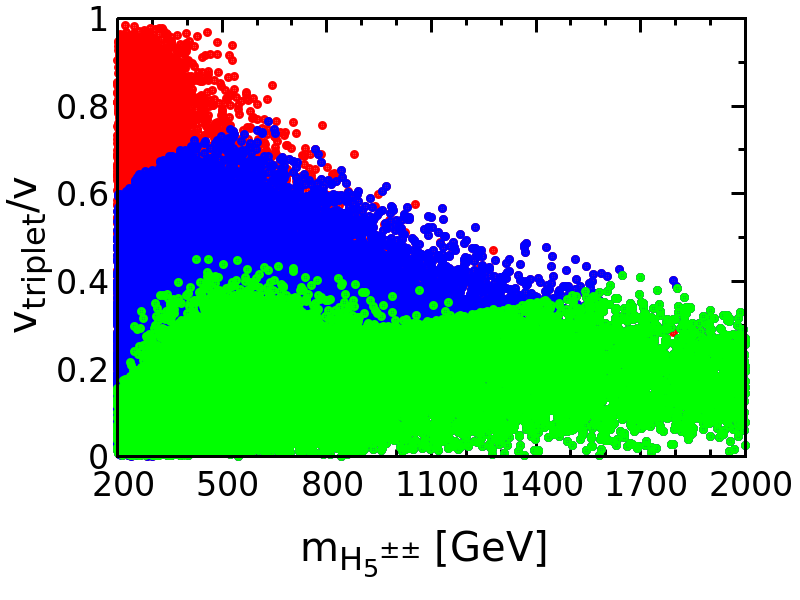}
          \caption{}
        \label{4b}
     \end{subfigure}
     \hspace{0.06\textwidth}
     \centering
     \begin{subfigure}[t']{0.25\textwidth}
         \centering
          \includegraphics[width=1.3\textwidth]{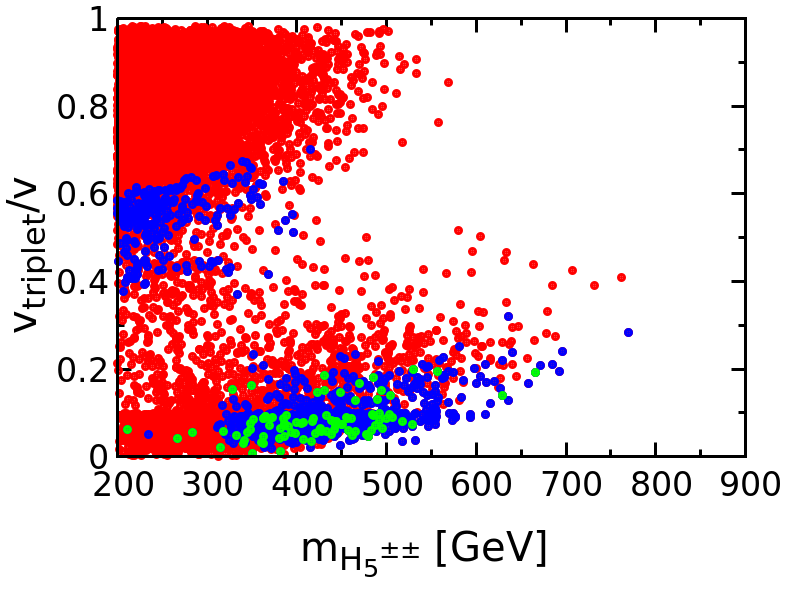}
          \caption{}
        \label{4c}
     \end{subfigure}
     \caption{Constraints on the $m_{H_5^{\pm\pm}} - (\frac{v_{triplet}}{v})$ plane for $v_\chi \ne v_\xi$. 1(a),(b) and (c) corresponds respectively to the situation of single, double and triple channel decay of $H_5^{++}$. Red, black and blue regions are excluded by indirect constraints, DY search of $H_5^{++}$ and VBF of $H_5^{++}$ respectively. Green regions are allowed by all constraints.}
     \label{4}
\end{figure}

\begin{figure}[!htb]
     \centering
     \begin{subfigure}[t']{0.25\textwidth}
         \centering
         \includegraphics[width=1.3\textwidth]{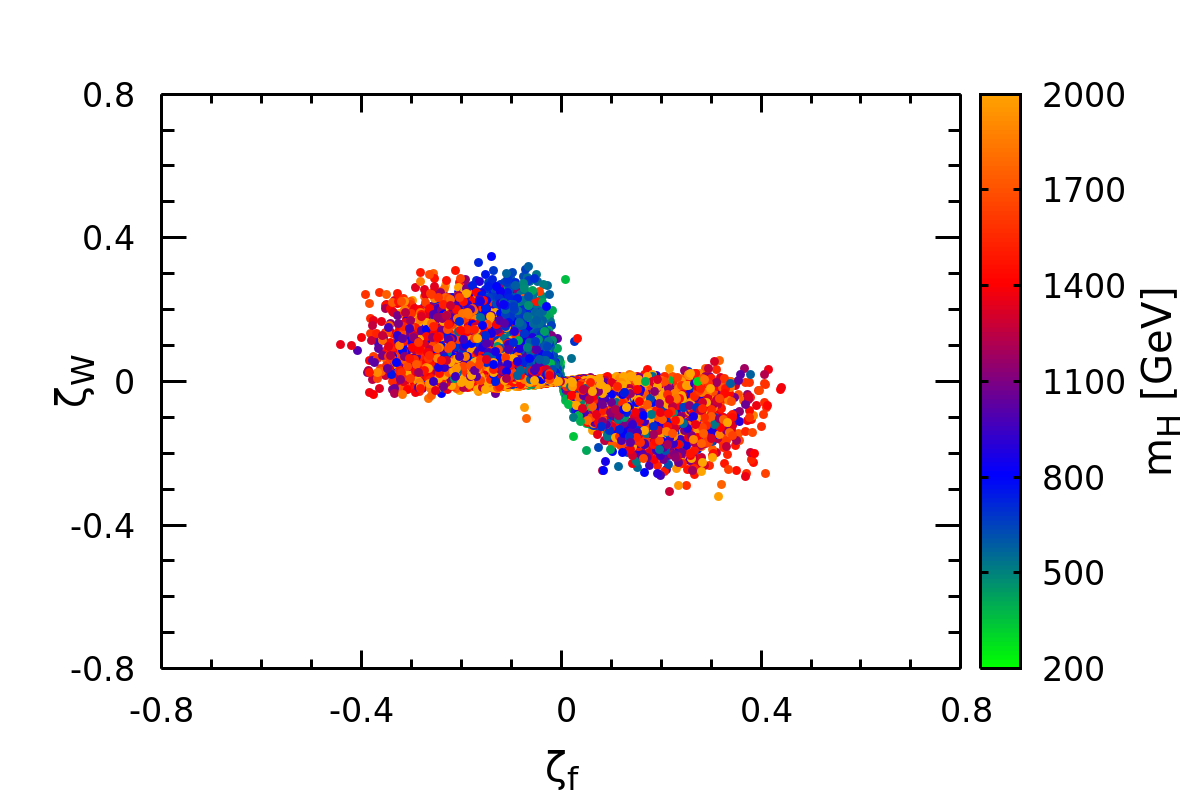}
         \caption{}
         \label{5a}
     \end{subfigure}
     \hspace{0.06\textwidth}
     \begin{subfigure}[h]{0.25\textwidth}
         \centering
         \includegraphics[width=1.3\textwidth]{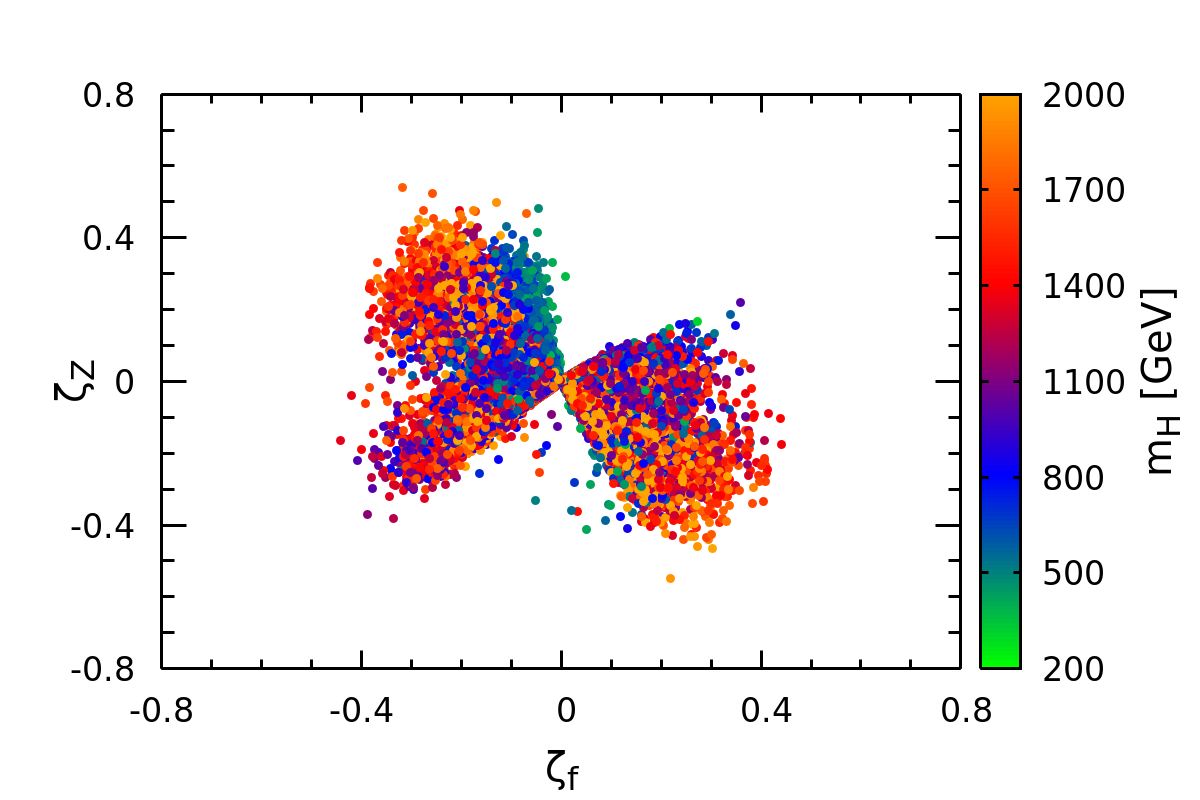}
         \caption{}
         \label{5b}
     \end{subfigure}
     \hspace{0.06\textwidth}
     \begin{subfigure}[h]{0.25\textwidth}
         \centering
         \includegraphics[width=1.3\textwidth]{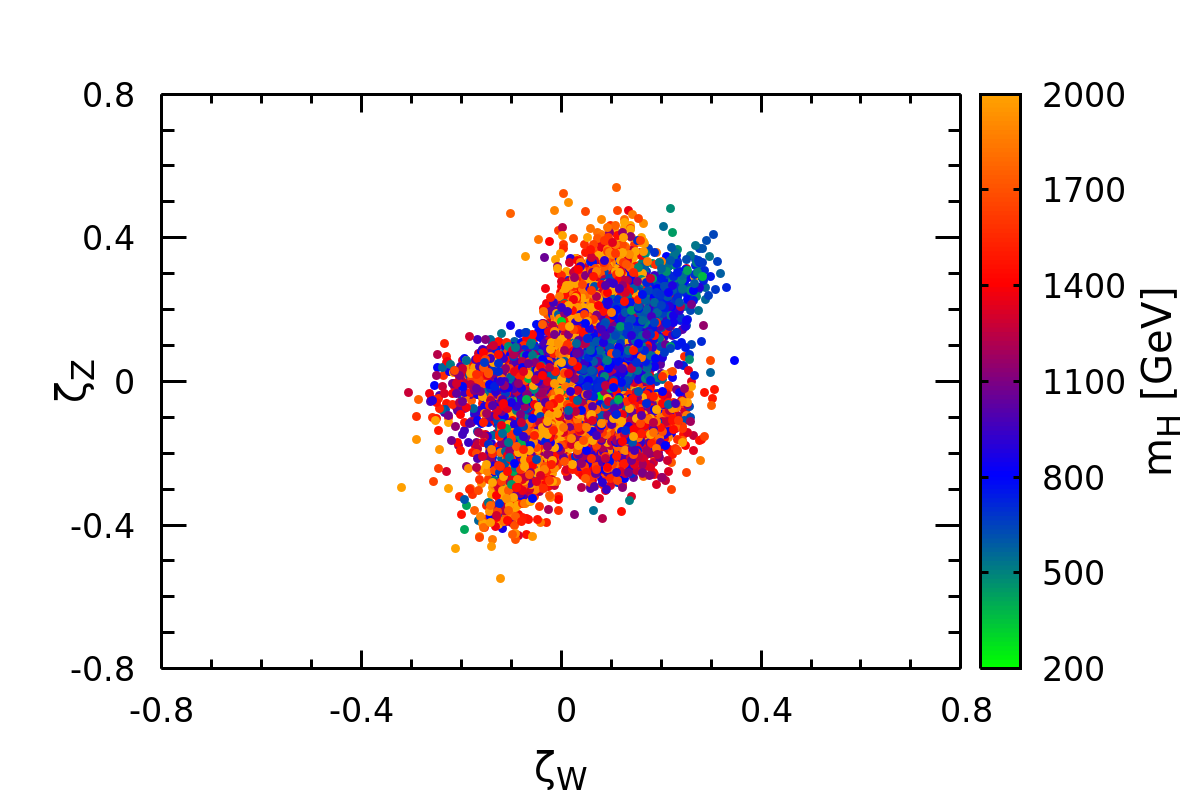}
         \caption{}
         \label{5c}
     \end{subfigure}
     \centering
     \begin{subfigure}[h]{0.25\textwidth}
         \centering
         \includegraphics[width=1.3\textwidth]{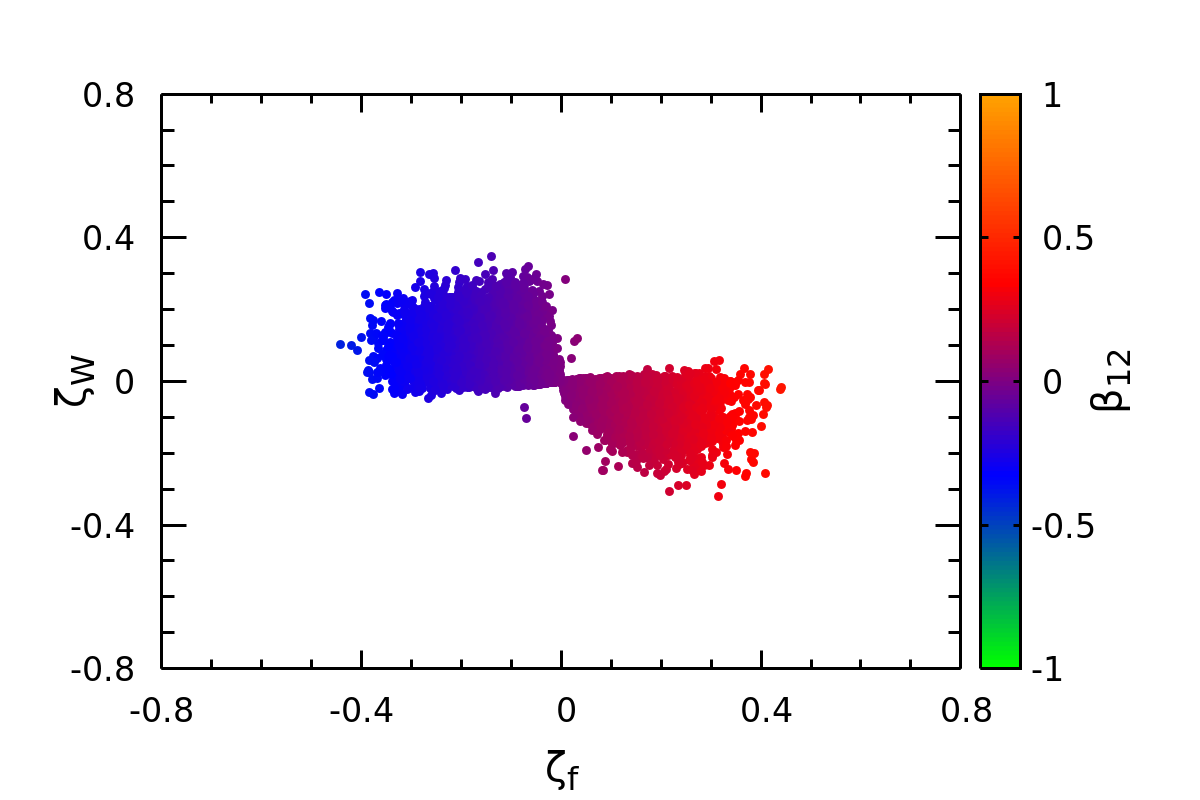}
         \caption{}
         \label{5d}
     \end{subfigure}
     \hspace{0.06\textwidth}
     \begin{subfigure}[h]{0.25\textwidth}
         \centering
         \includegraphics[width=1.3\textwidth]{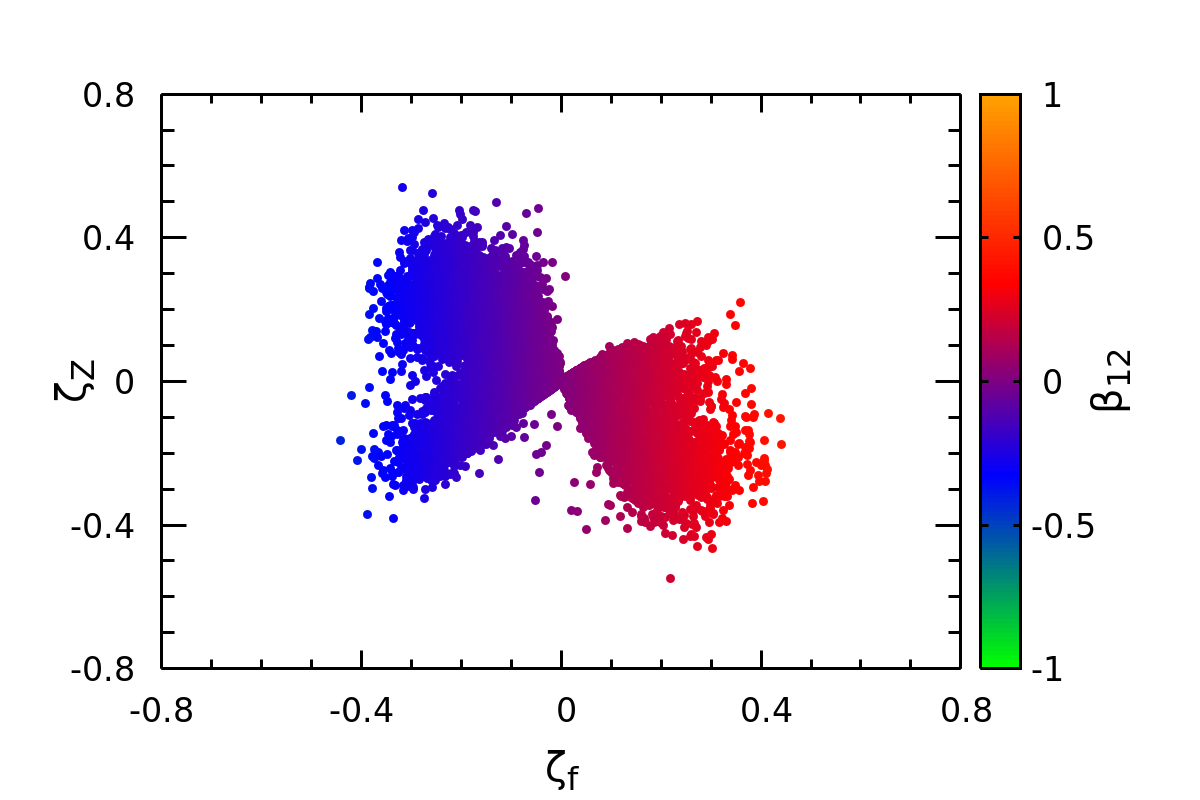}
         \caption{}
         \label{5e}
     \end{subfigure}
     \hspace{0.06\textwidth}
     \begin{subfigure}[h]{0.25\textwidth}
         \centering
         \includegraphics[width=1.3\textwidth]{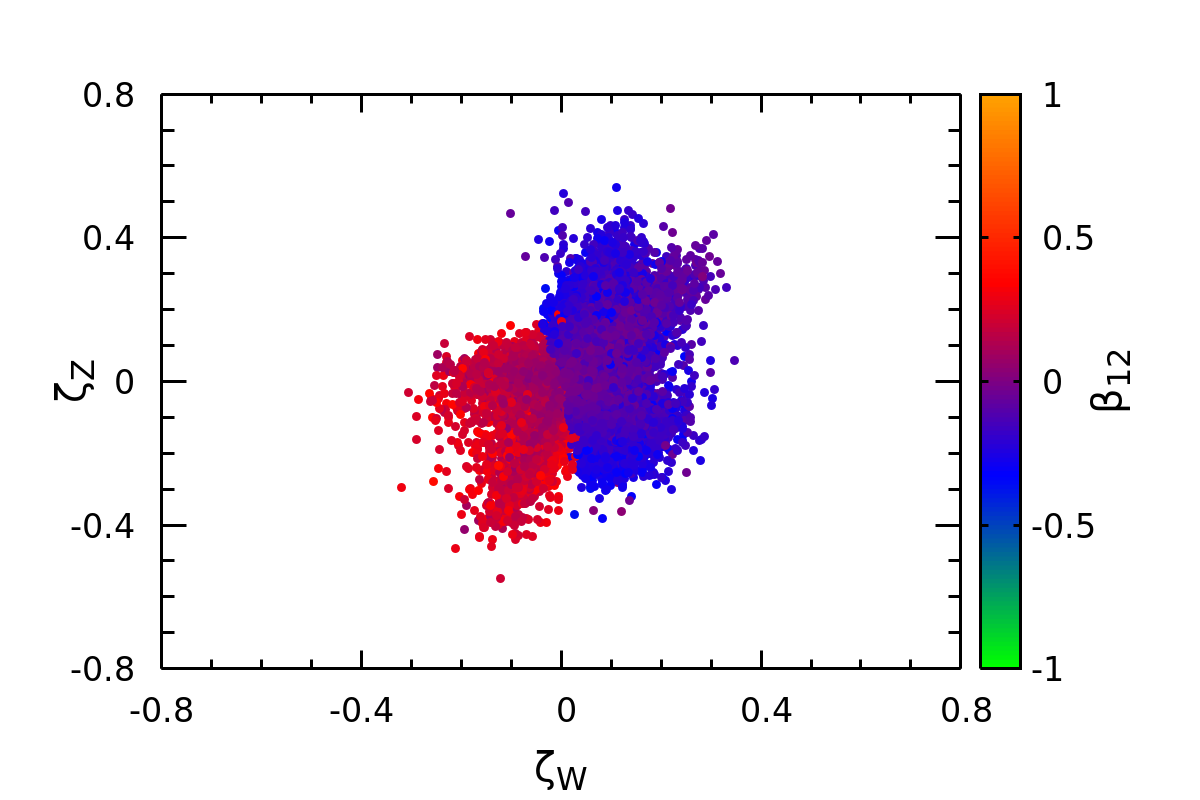}
         \caption{}
         \label{5f}
     \end{subfigure}
     \caption{Correlated modification factors in coupling strengths for the neutral scalar H to $W$, $Z$ and fermion pairs with respect to the corresponding couplings of the SM Higgs for single channel decay of $H_5^{++}$ for $v_\chi \ne v_\xi$. The colour axis in (a),(b) and (c) correspond to $m_H $, the mass of H while in (d),(e) and (f) it correspond to $\beta_{12}$, the $SU(2)_L$ doublet content of H.} 
     \label{5}
\end{figure}

\begin{figure}[!htb]
     \centering
     \begin{subfigure}[t']{0.25\textwidth}
         \centering
         \includegraphics[width=1.3\textwidth]{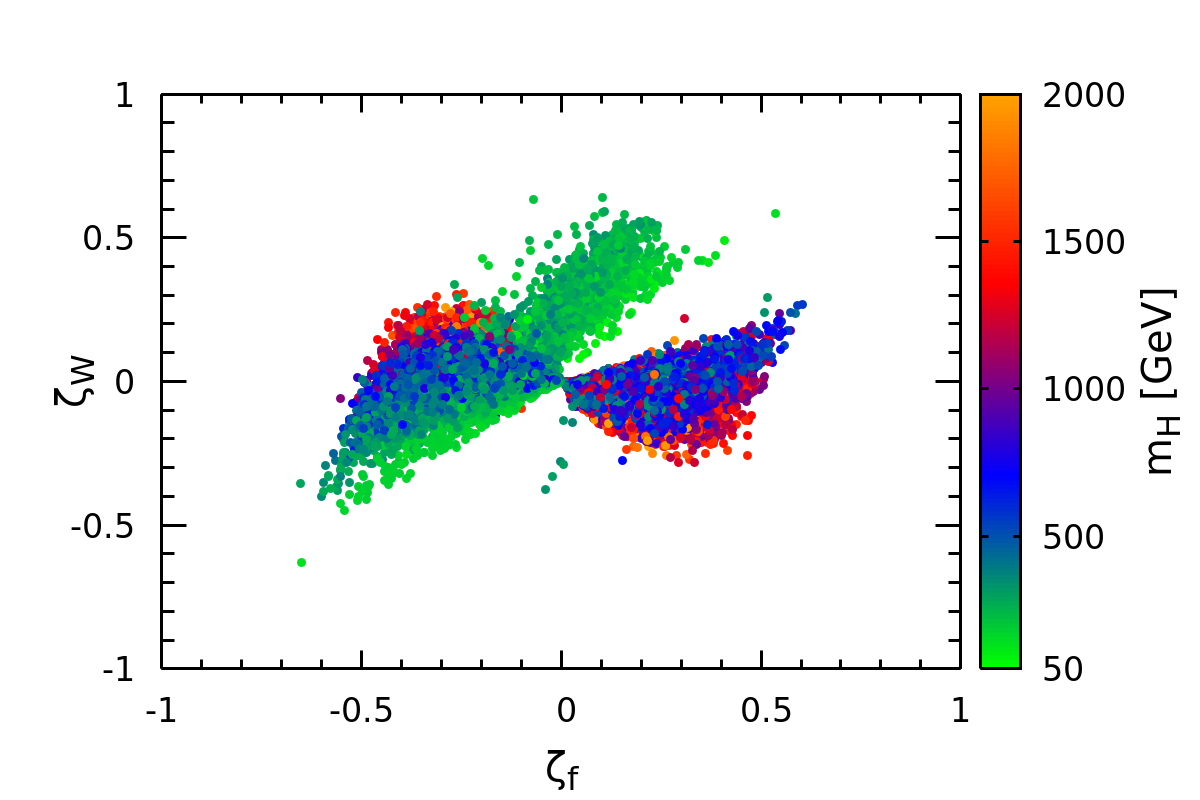}
         \caption{}
         \label{6a}
     \end{subfigure}
     \hspace{0.06\textwidth}
     \begin{subfigure}[h]{0.25\textwidth}
         \centering
         \includegraphics[width=1.3\textwidth]{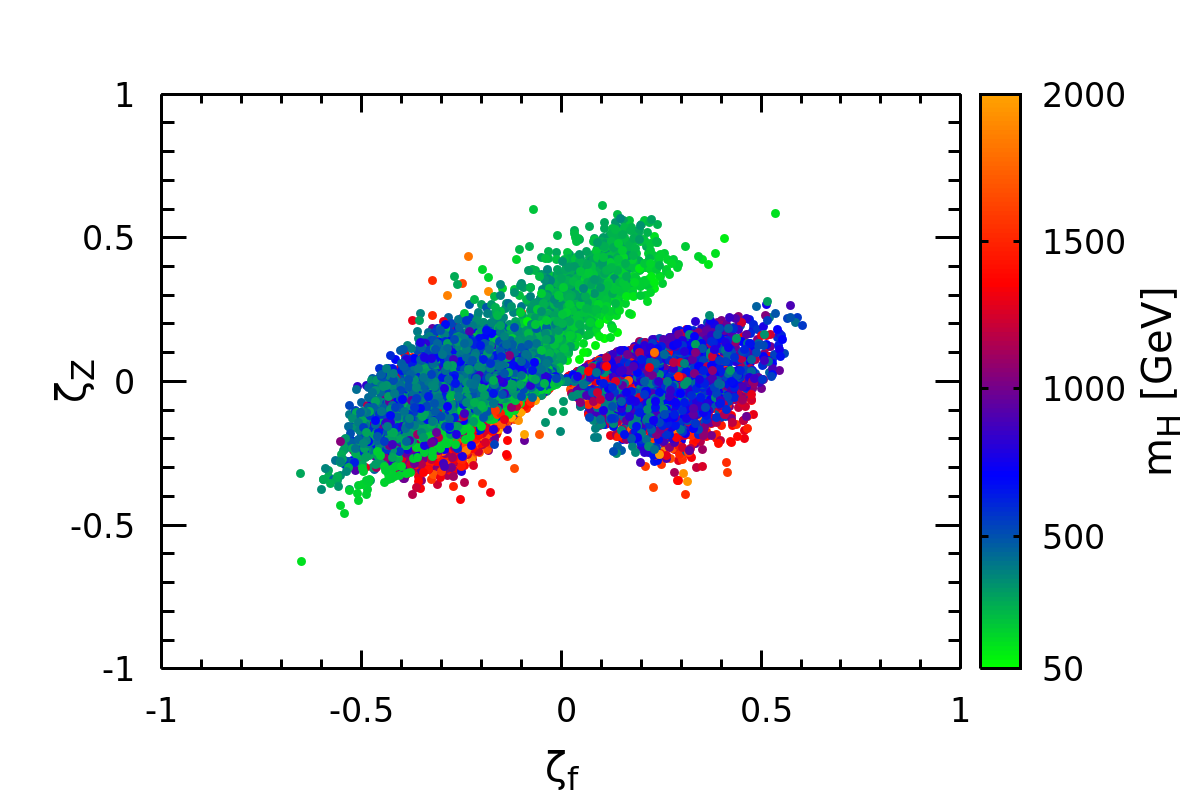}
         \caption{}
         \label{6b}
     \end{subfigure}
     \hspace{0.06\textwidth}
     \begin{subfigure}[h]{0.25\textwidth}
         \centering
         \includegraphics[width=1.3\textwidth]{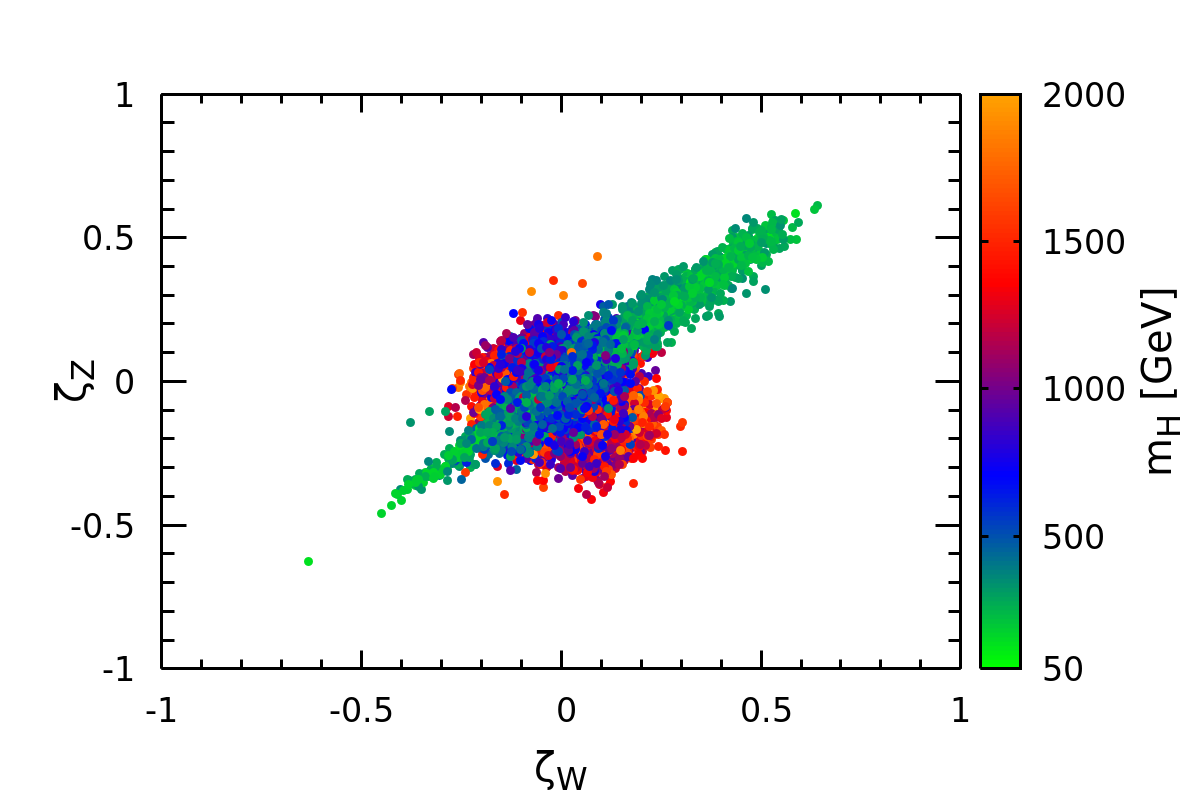}
         \caption{}
         \label{6c}
     \end{subfigure}
     \centering
     \begin{subfigure}[h]{0.25\textwidth}
         \centering
         \includegraphics[width=1.3\textwidth]{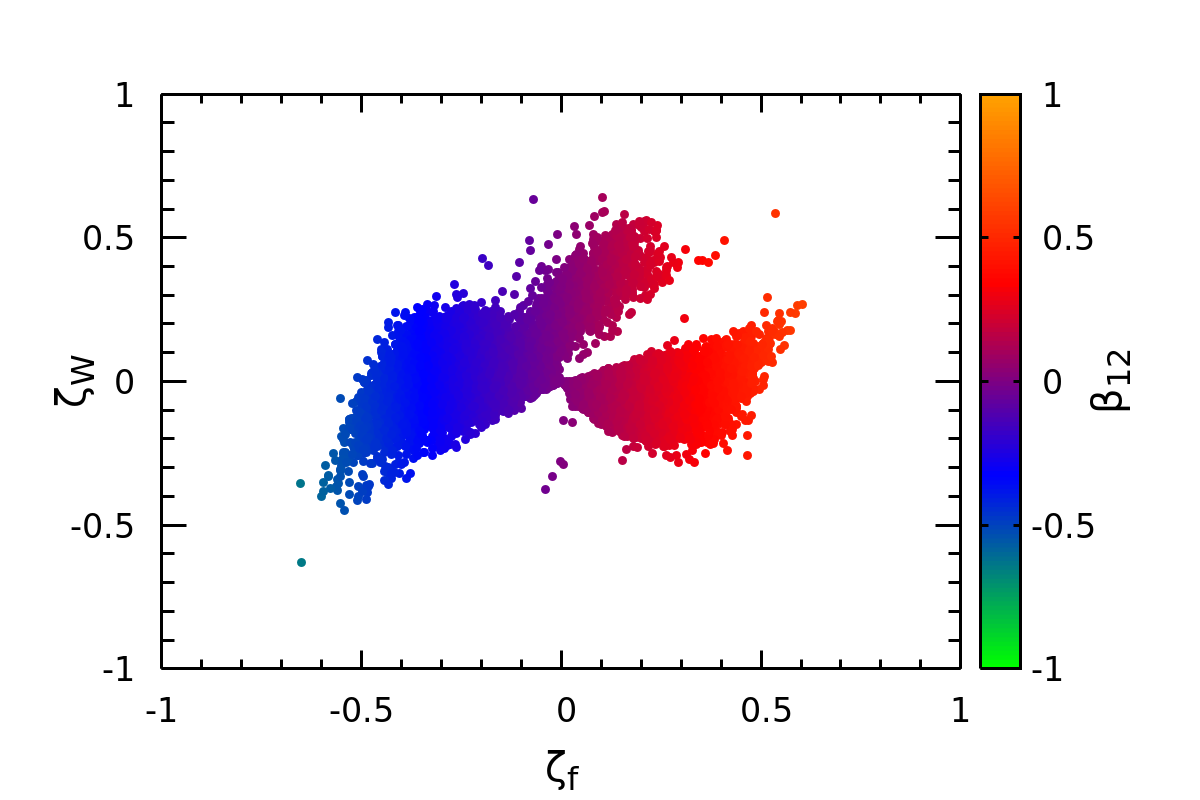}
         \caption{}
         \label{6d}
     \end{subfigure}
     \hspace{0.06\textwidth}
     \begin{subfigure}[h]{0.25\textwidth}
         \centering
         \includegraphics[width=1.3\textwidth]{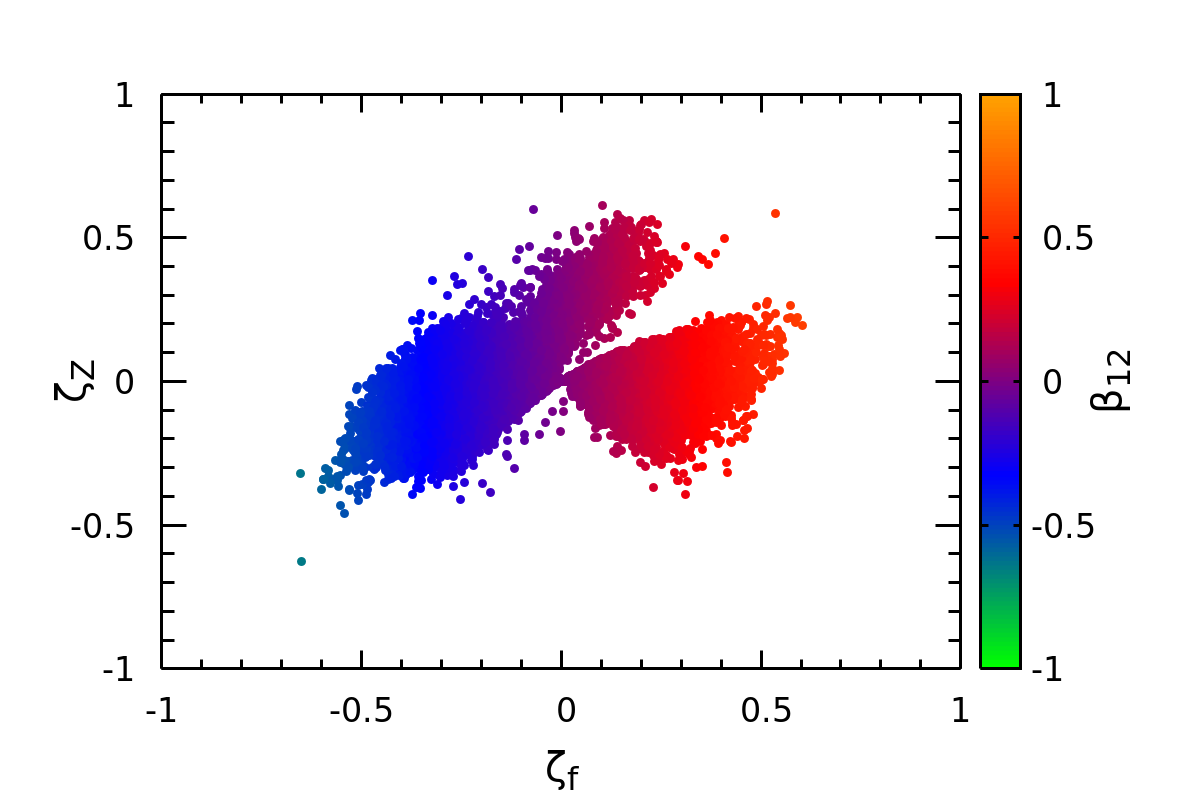}
         \caption{}
         \label{6e}
     \end{subfigure}
     \hspace{0.06\textwidth}
     \begin{subfigure}[h]{0.25\textwidth}
         \centering
         \includegraphics[width=1.3\textwidth]{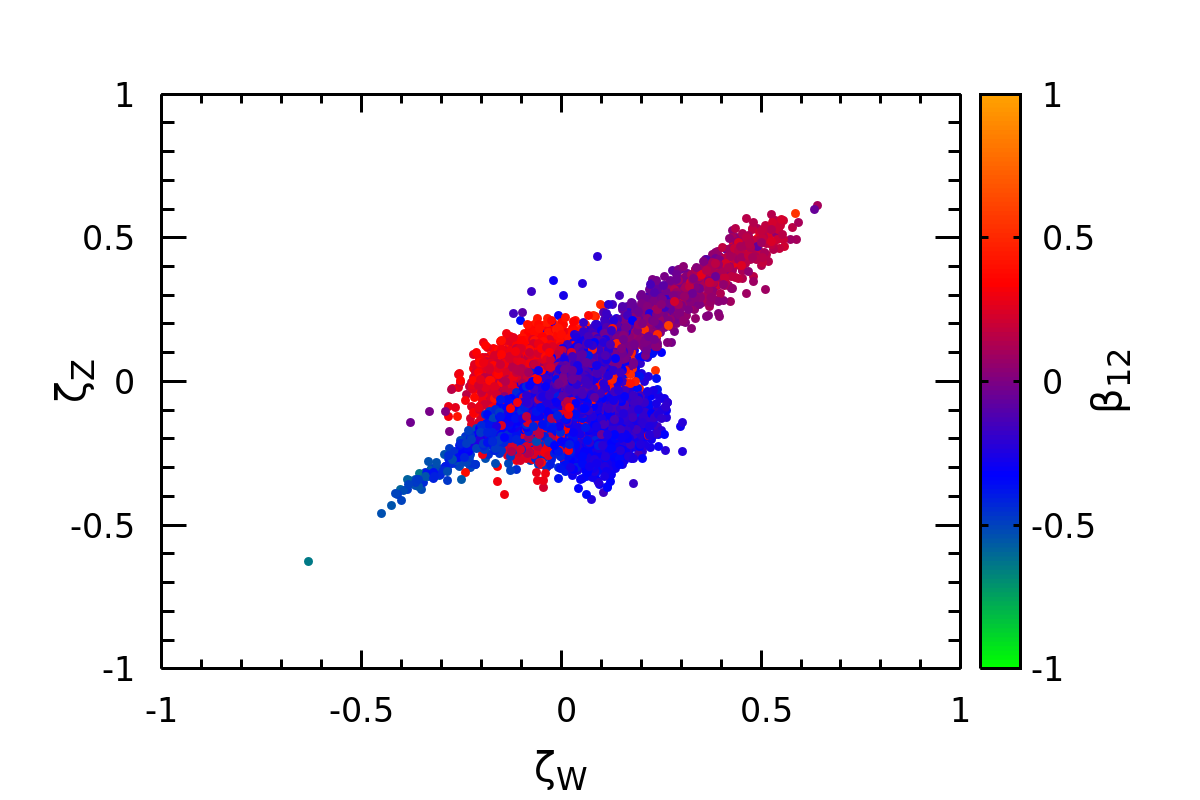}
         \caption{}
         \label{6f}
     \end{subfigure}
     \caption{Correlated modification factors in coupling strengths for the neutral scalar H to $W$, $Z$ and fermion pairs with respect to the corresponding couplings of the SM Higgs for double channel decay of $H_5^{++}$ for $v_\chi \ne v_\xi$. The colour axis in (a),(b) and (c) correspond to $m_H$, the mass of H while in (d),(e) and (f) it correspond to $\beta_{12}$, the $SU(2)_L$ doublet content of H.}
     \label{6}
\end{figure}

\begin{figure}[!htb]
     \centering
     \begin{subfigure}[h]{0.25\textwidth}
         \centering
         \includegraphics[width=1.3\textwidth]{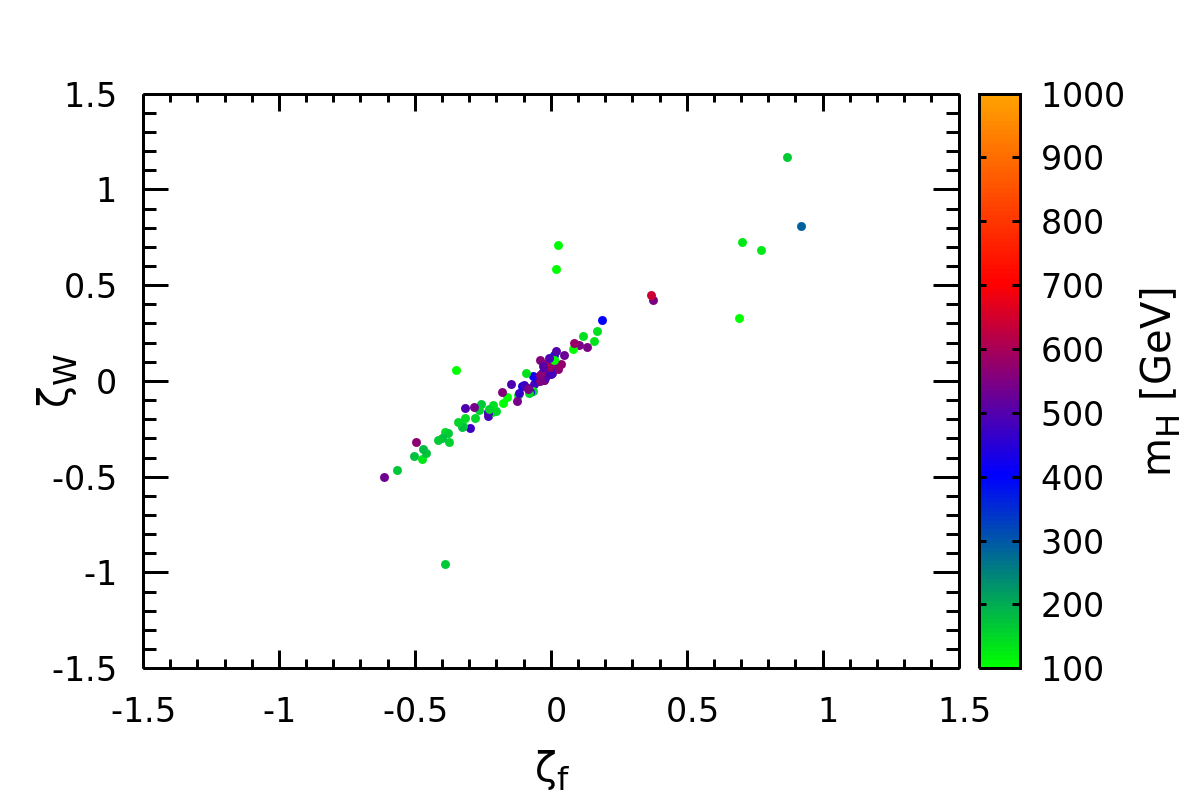}
         \caption{}
         \label{7a}
     \end{subfigure}
     \hspace{0.06\textwidth}
     \begin{subfigure}[h]{0.25\textwidth}
         \centering
         \includegraphics[width=1.3\textwidth]{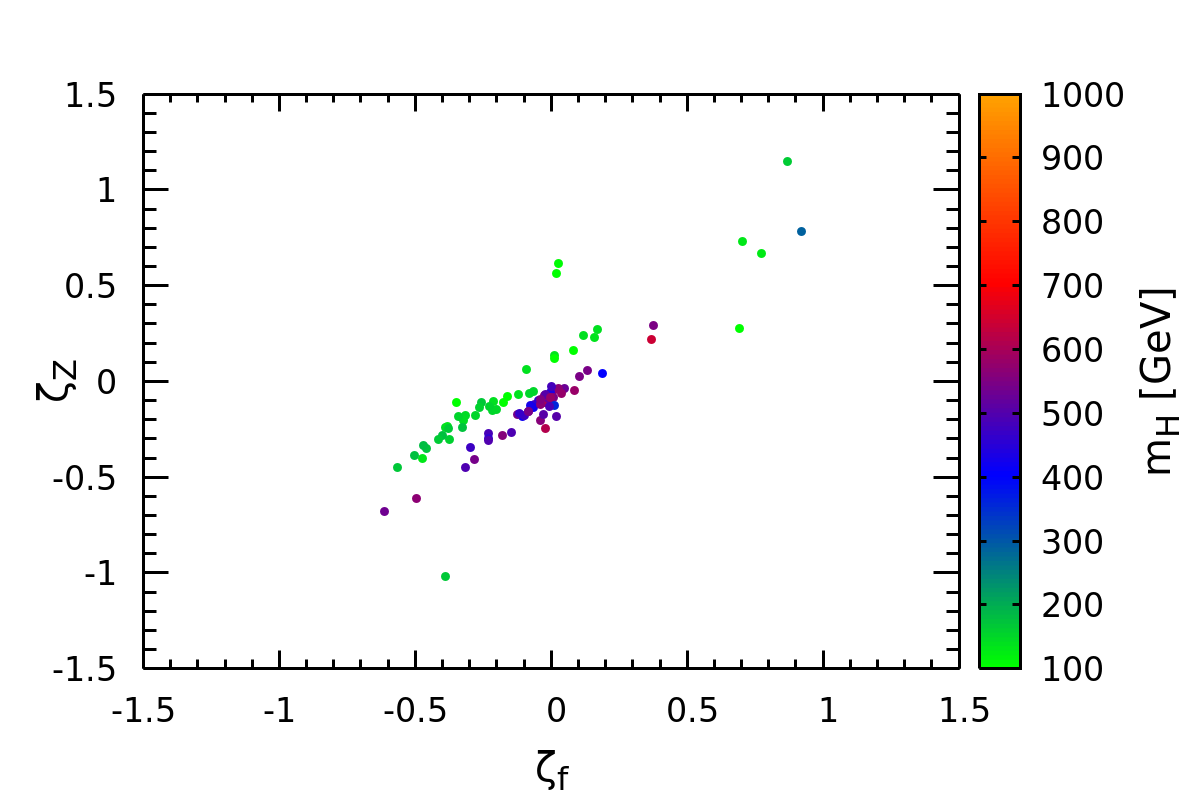}
         \caption{}
         \label{7b}
     \end{subfigure}
     \hspace{0.06\textwidth}
     \begin{subfigure}[h]{0.25\textwidth}
         \centering
         \includegraphics[width=1.3\textwidth]{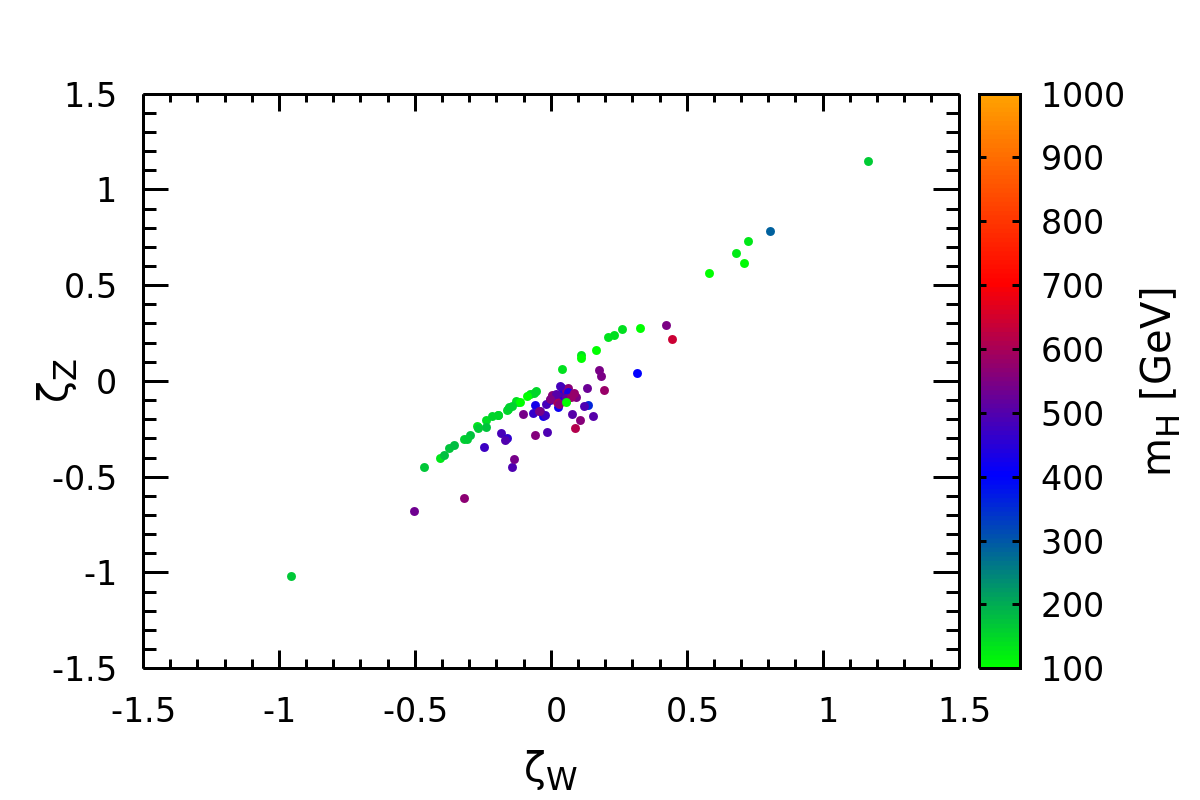}
         \caption{}
         \label{7c}
     \end{subfigure}
     \centering
     \begin{subfigure}[h]{0.25\textwidth}
         \centering
         \includegraphics[width=1.3\textwidth]{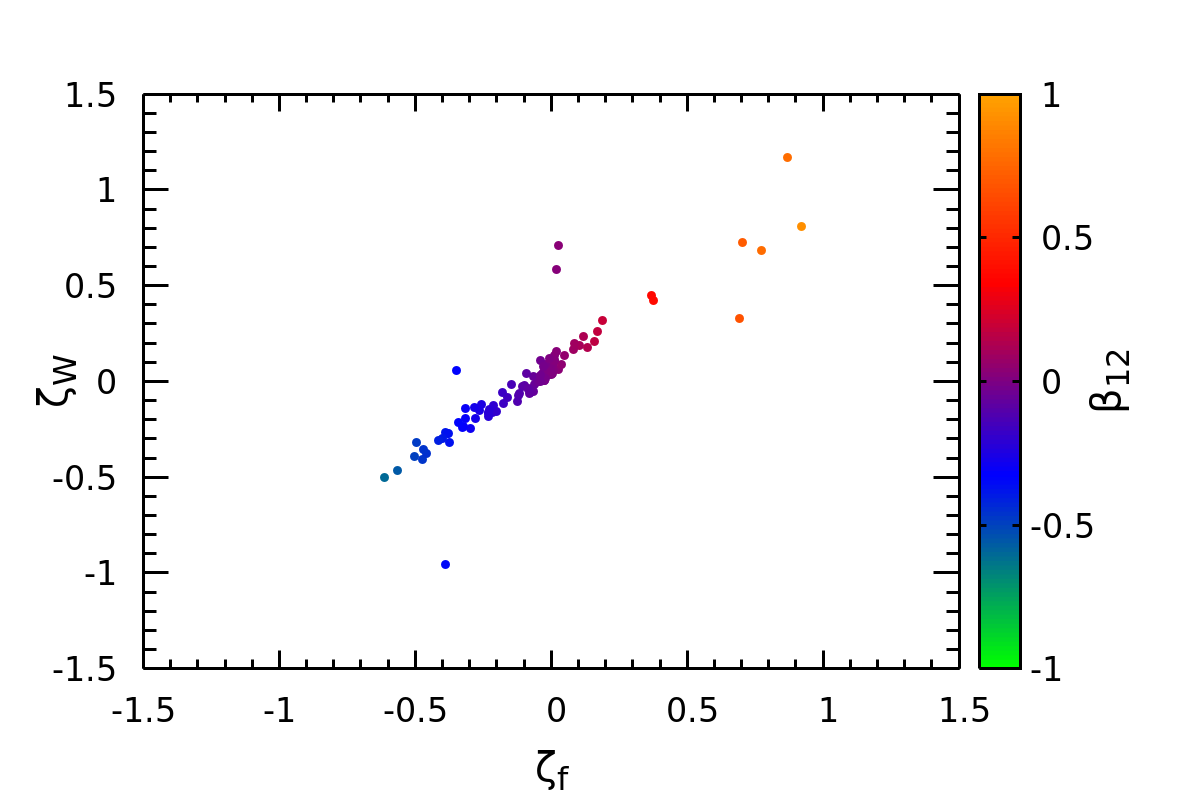}
         \caption{}
         \label{7d}
     \end{subfigure}
     \hspace{0.06\textwidth}
     \begin{subfigure}[h]{0.25\textwidth}
         \centering
         \includegraphics[width=1.3\textwidth]{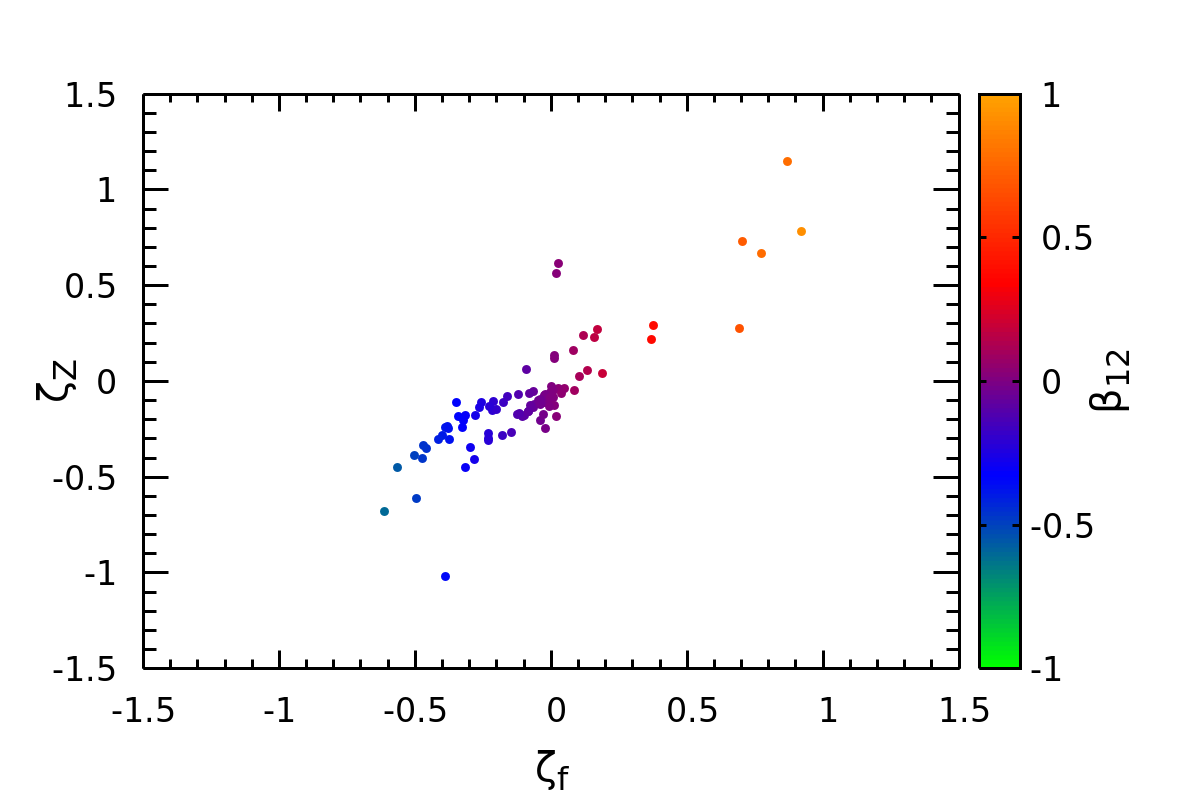}
         \caption{}
         \label{7e}
     \end{subfigure}
     \hspace{0.06\textwidth}
     \begin{subfigure}[h]{0.25\textwidth}
         \centering
         \includegraphics[width=1.3\textwidth]{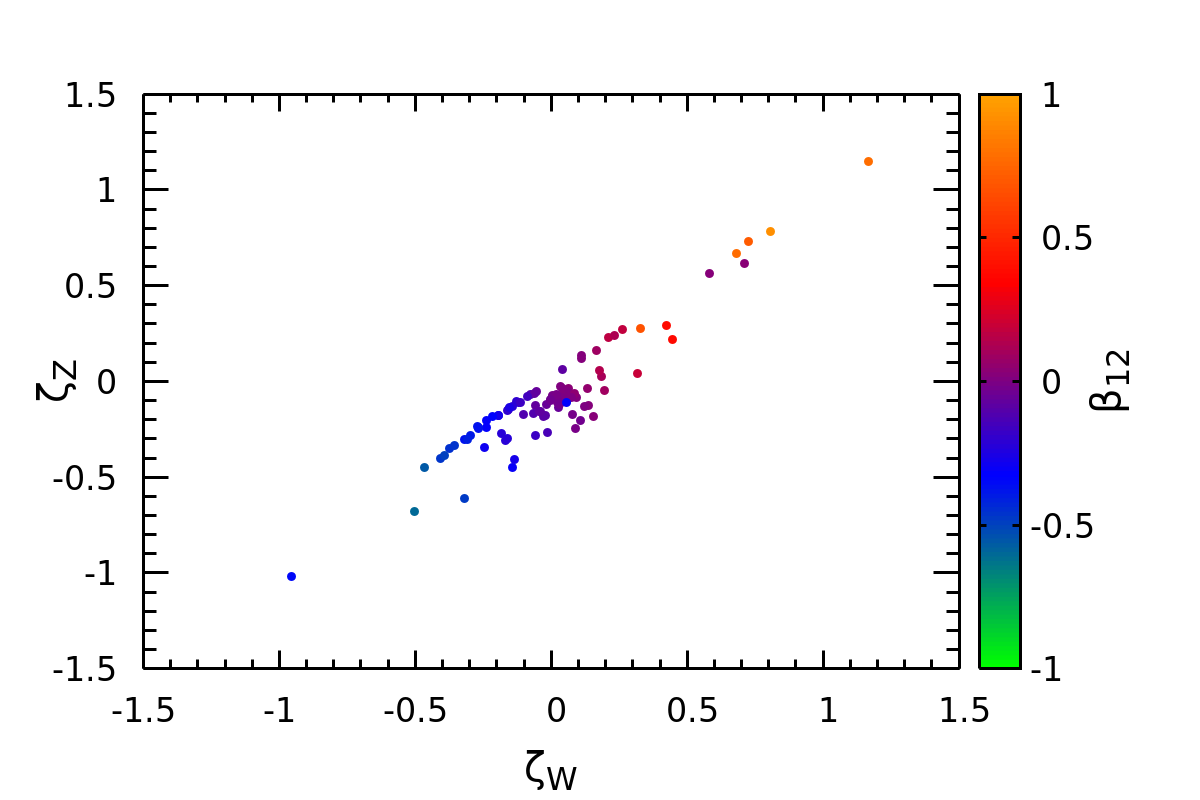}
         \caption{}
         \label{7f}
     \end{subfigure}
     \caption{Correlated modification factors in coupling strengths for the neutral scalar H to $W$, $Z$ and fermion pairs with respect to the corresponding couplings of the SM Higgs for triple channel decay of $H_5^{++}$ for $v_\chi \ne v_\xi$. The colour axis in (a),(b) and (c) correspond to $m_H$, the mass of H while in (d),(e) and (f) it correspond to $\beta_{12}$, the $SU(2)_L$ doublet content of H.}
     \label{7}
\end{figure}

\begin{figure}[!htb]
     \centering
     \begin{subfigure}[h]{0.25\textwidth}
         \centering
         \includegraphics[width=1.3\textwidth]{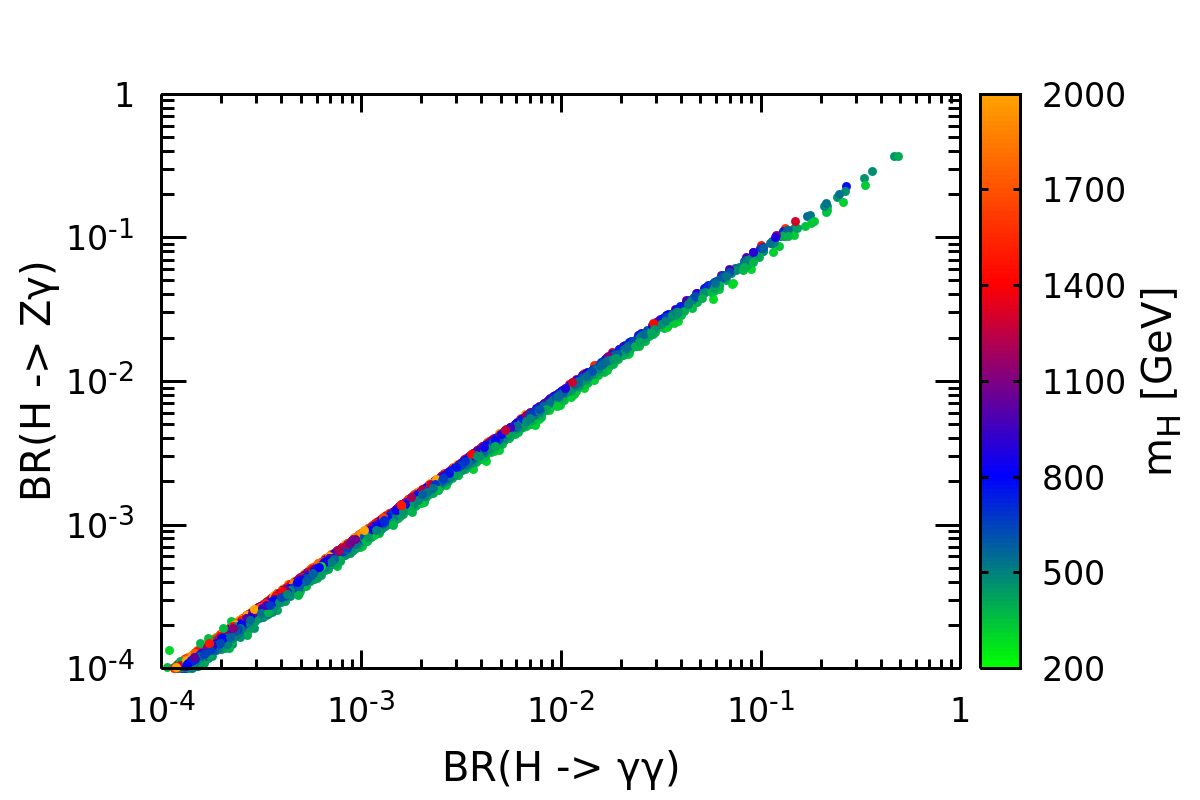}
         \caption{}
         \label{8a}
     \end{subfigure}
     \hspace{0.06\textwidth}
     \begin{subfigure}[h]{0.25\textwidth}
         \centering
         \includegraphics[width=1.3\textwidth]{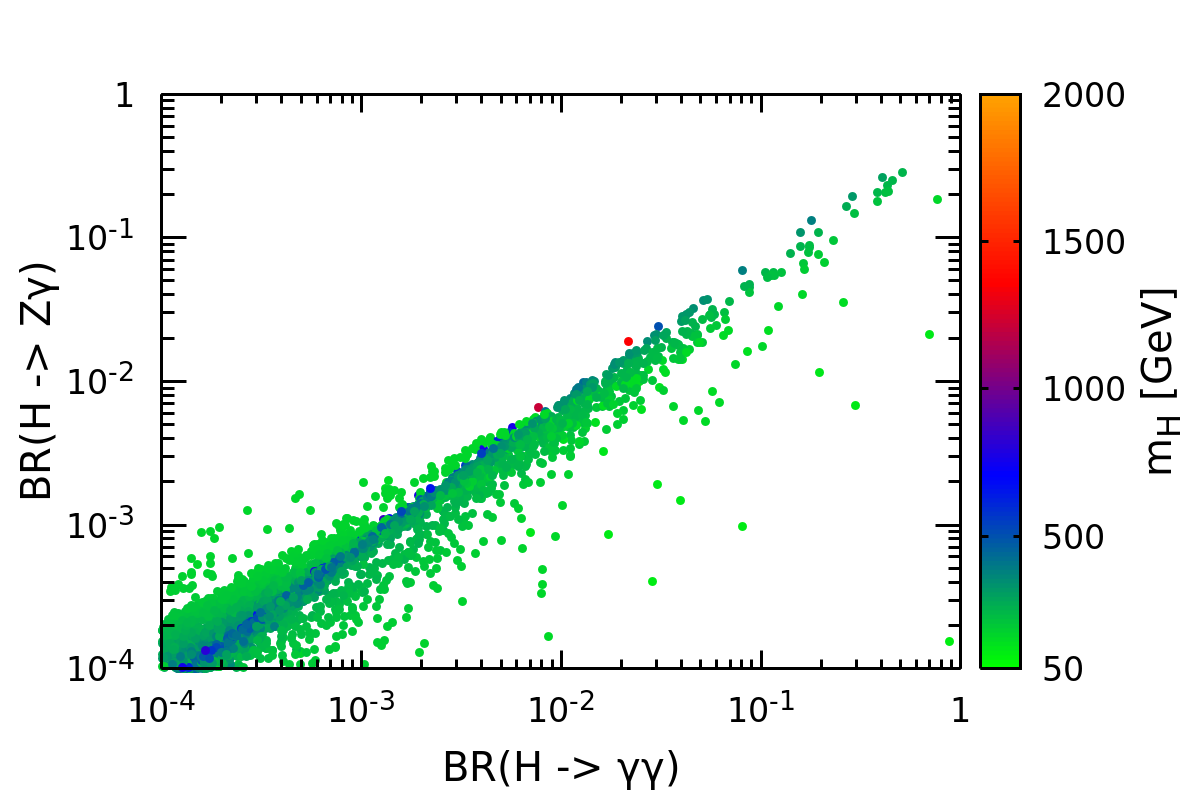}
         \caption{}
         \label{8b}
     \end{subfigure}
     \hspace{0.06\textwidth}
     \begin{subfigure}[h]{0.25\textwidth}
         \centering
         \includegraphics[width=1.3\textwidth]{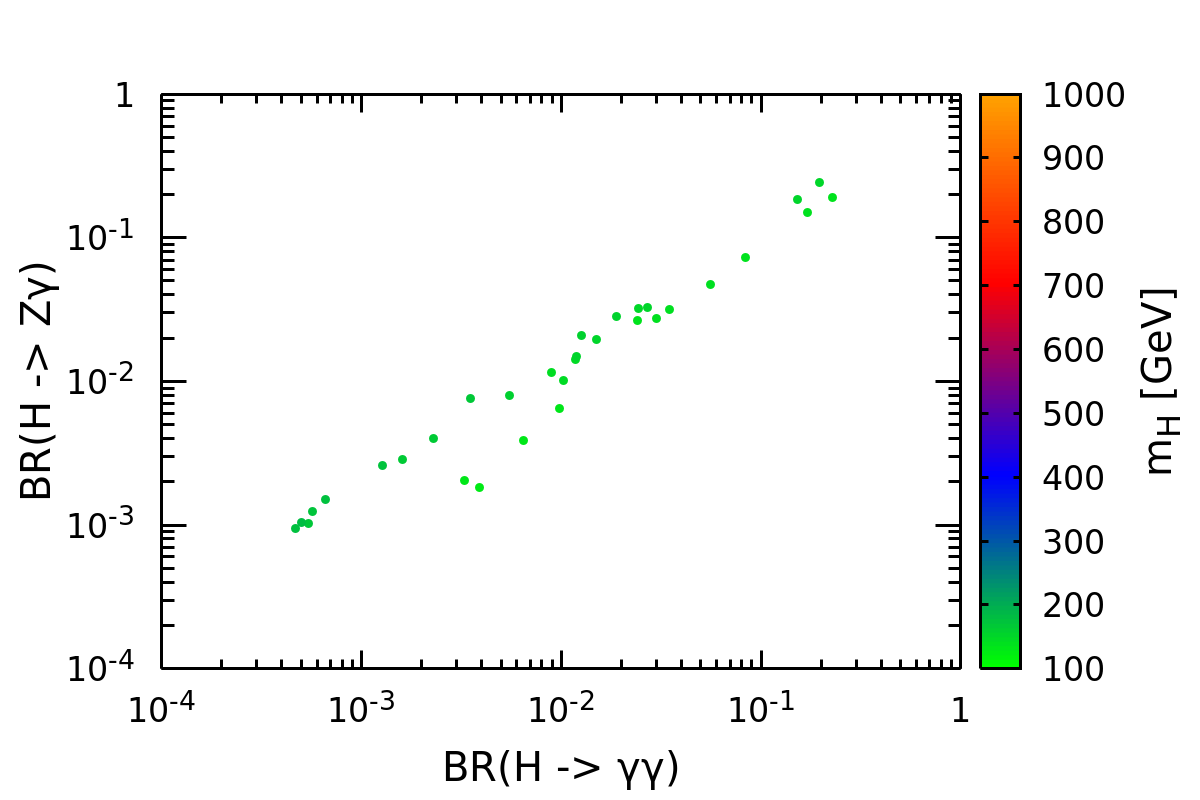}
         \caption{}
         \label{8c}
     \end{subfigure}
     \centering
     \begin{subfigure}[h]{0.25\textwidth}
         \centering
         \includegraphics[width=1.3\textwidth]{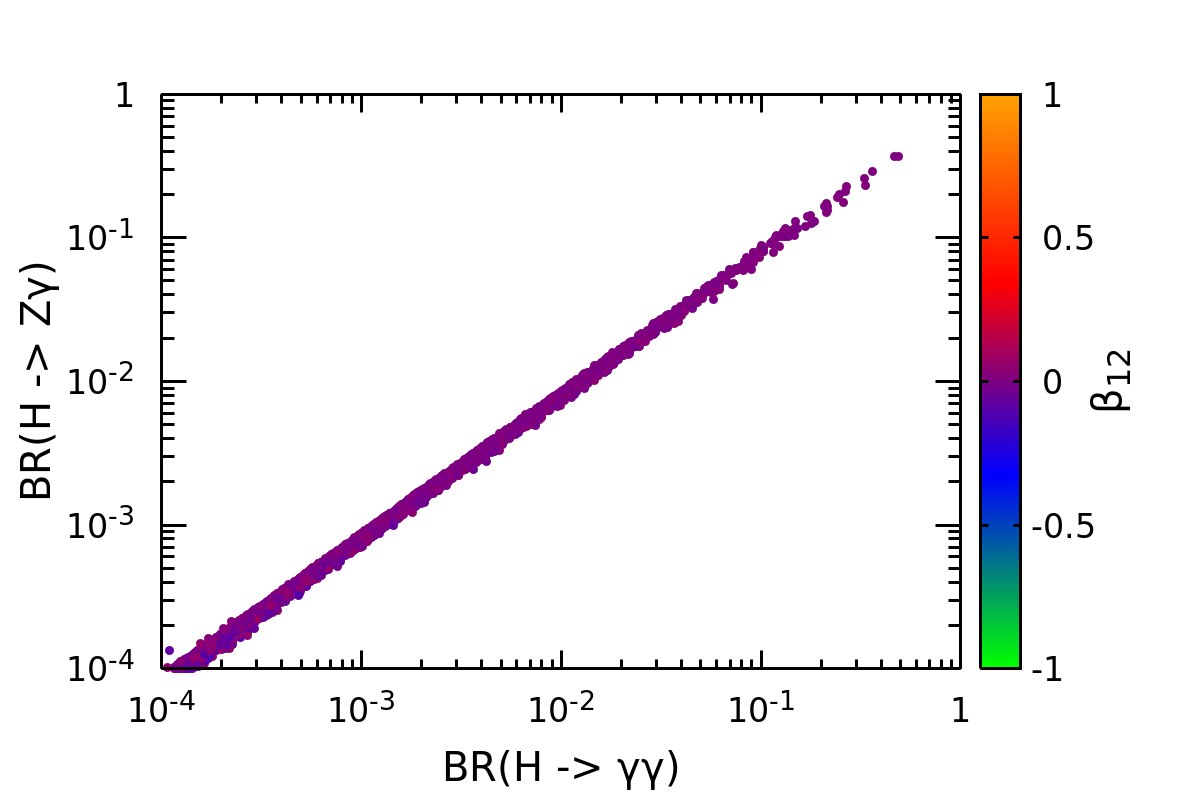}
         \caption{}
         \label{8d}
     \end{subfigure}
     \hspace{0.06\textwidth}
     \begin{subfigure}[h]{0.25\textwidth}
         \centering
         \includegraphics[width=1.3\textwidth]{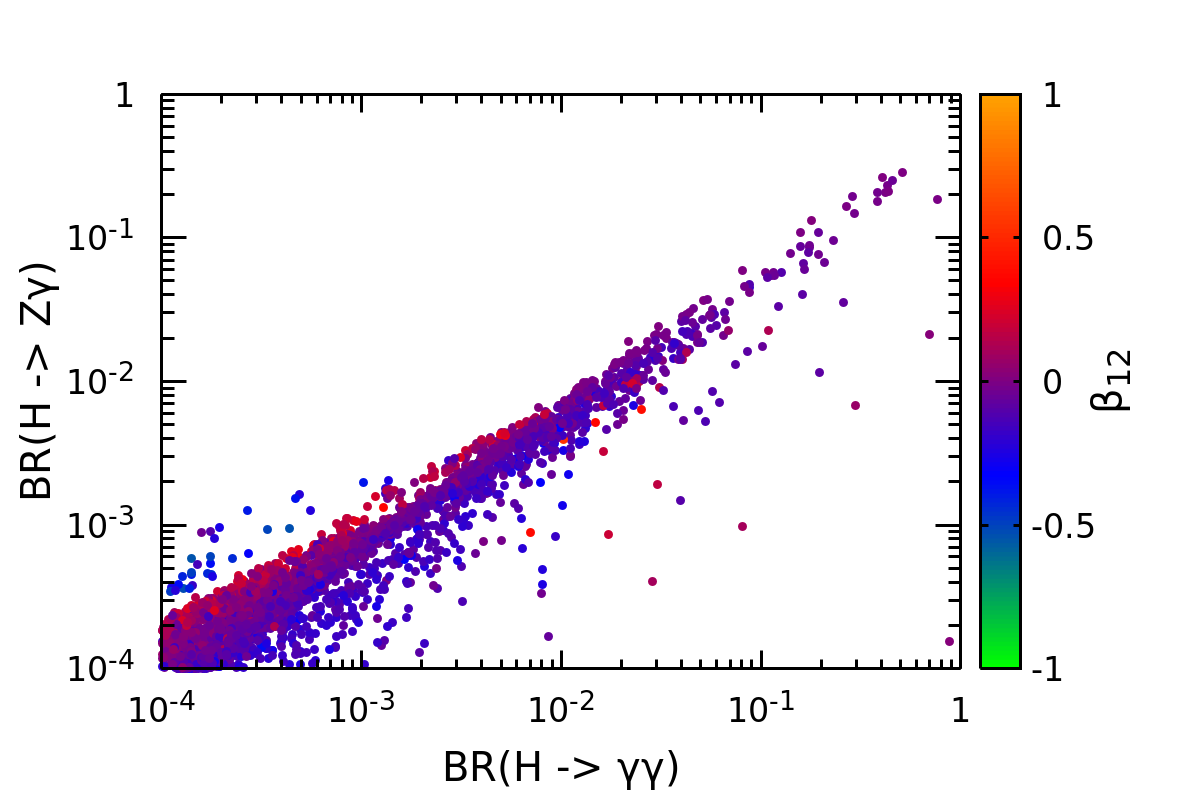}
         \caption{}
         \label{8d}
     \end{subfigure}
     \hspace{0.06\textwidth}
     \begin{subfigure}[h]{0.25\textwidth}
         \centering
         \includegraphics[width=1.3\textwidth]{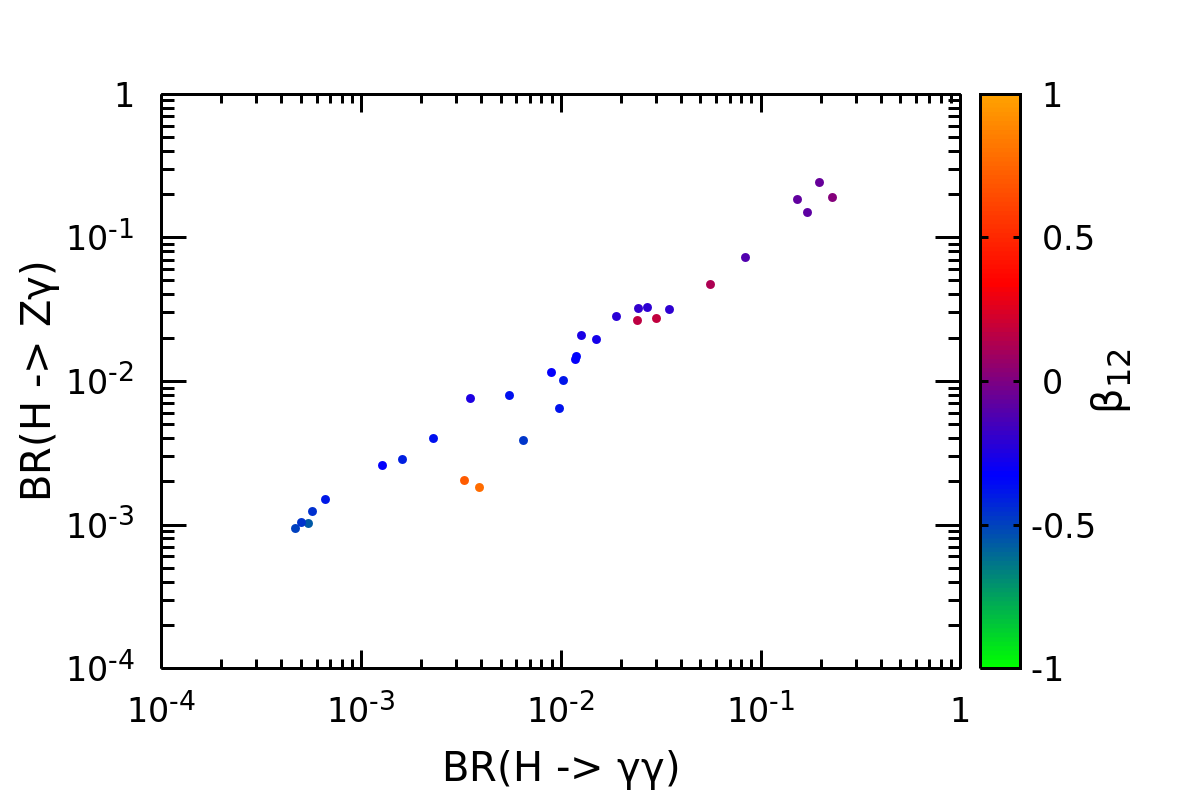}
         \caption{}
         \label{8f}
     \end{subfigure}
     \caption{Correlated branching ratios for $H \rightarrow Z \gamma$ and $H \rightarrow \gamma \gamma$ for single(a and d), double (b and e) and triple-channel (c and f) decay of $H_5^{++}$ for $v_\chi \ne v_\xi$. The colour axis in (a),(b) and (c) correspond to $m_H$, the mass of H while in (d),(e) and (f) it correspond to $\beta_{12}$, the $SU(2)_L$ doublet content of H.}
     \label{8}
\end{figure}

\begin{figure}[!htb]
     \centering
     \begin{subfigure}[h]{0.25\textwidth}
         \centering
         \includegraphics[width=1.3\textwidth]{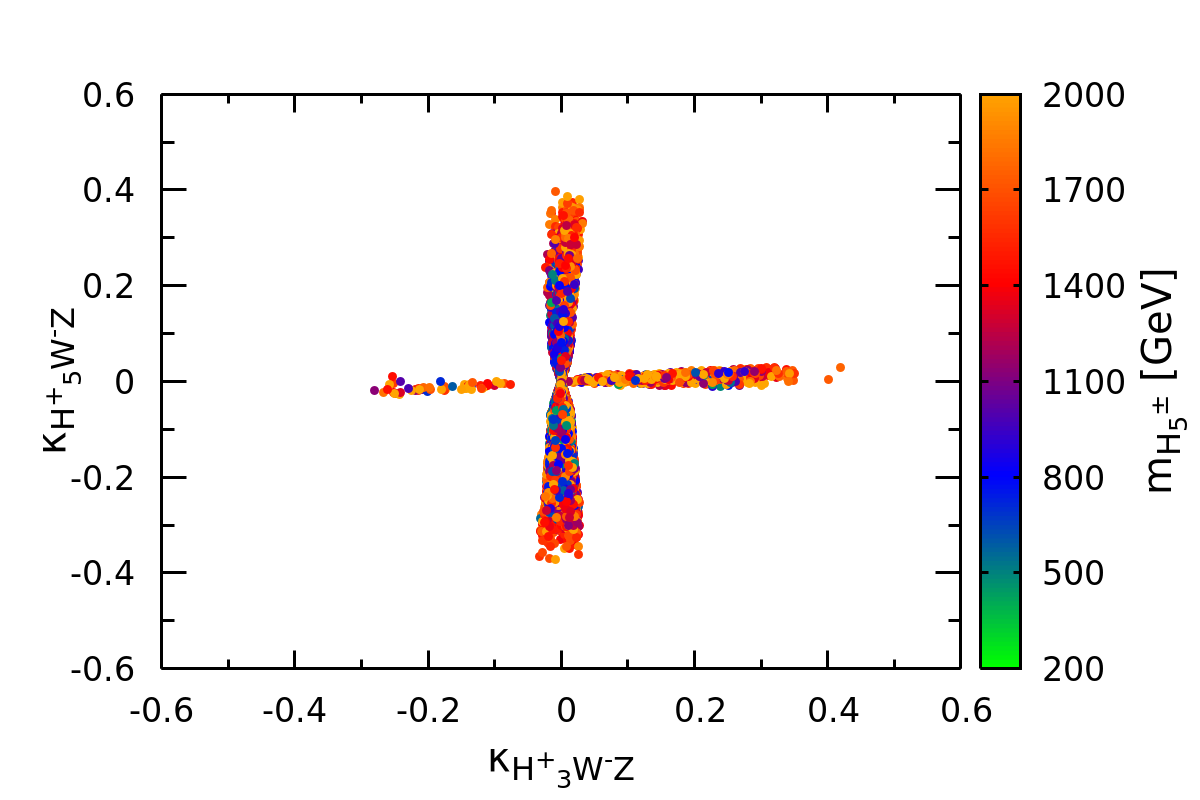}
         \caption{}
         \label{9a}
     \end{subfigure}
     \hspace{0.06\textwidth}
     \begin{subfigure}[h]{0.25\textwidth}
         \centering
         \includegraphics[width=1.3\textwidth]{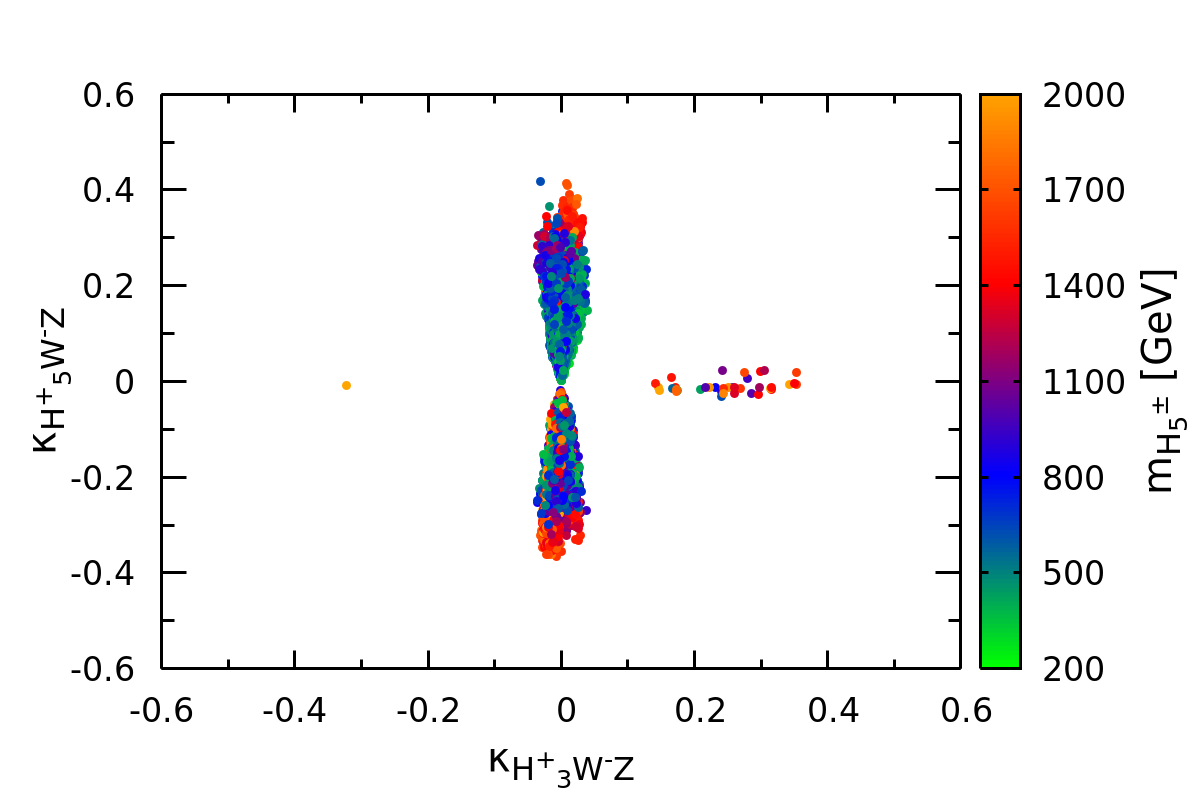}
         \caption{}
         \label{9b}
     \end{subfigure}
     \hspace{0.06\textwidth}
     \begin{subfigure}[h]{0.25\textwidth}
         \centering
         \includegraphics[width=1.3\textwidth]{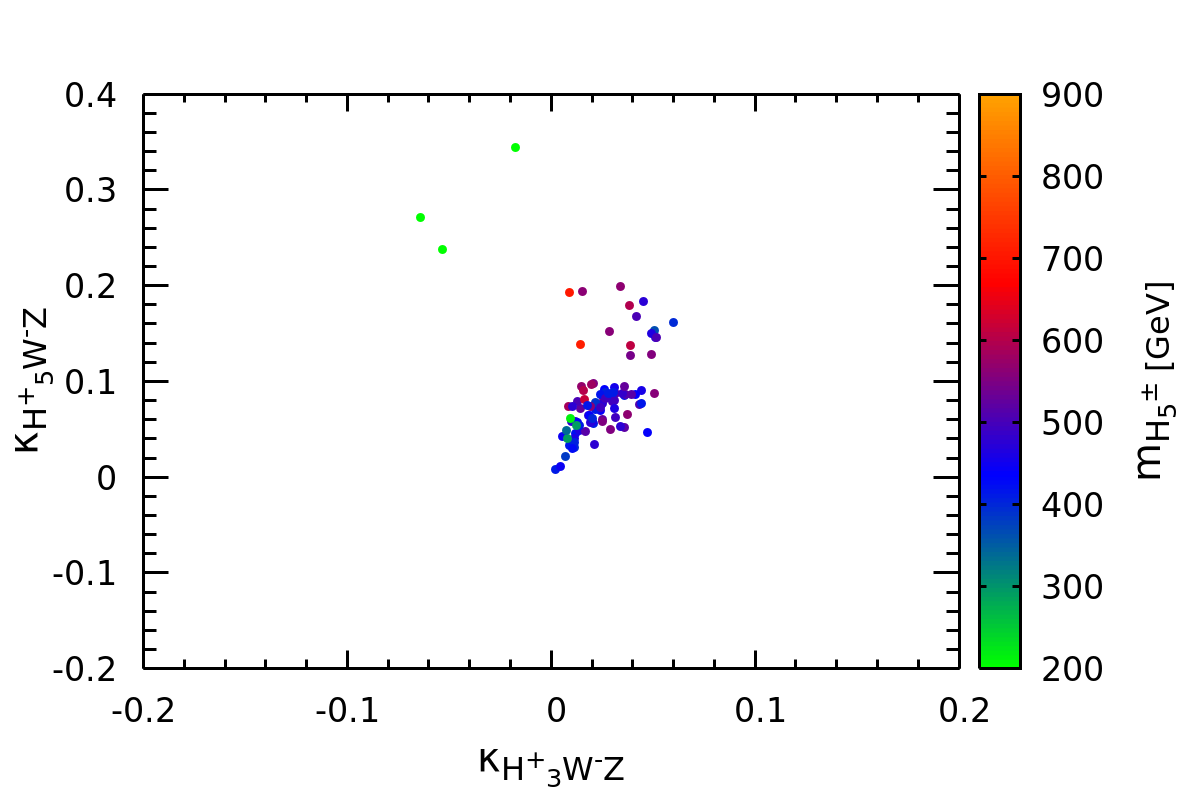}
         \caption{}
         \label{9c}
     \end{subfigure}
     \caption{The correlated vertex strength modification factors of $H_5^+W^-Z$ and $H_3^+W^-Z$ for $v_\chi \ne v_\xi$ with respect to the corresponding $H_5^+W^-Z$ coupling strength in custodial symmetric case, as in eq. \eqref{hwzcs} with $s_H = 1$. The colour axis represents $m_{H_5^\pm}$, the mass of the singly charged scalar which corresponds to $H_5^+$ in custodial symmetric limit. Again (a),(b),(c)  correspond respectively to single, double and triple channel decays of $H_5^{++}$.} 
     \label{9}
\end{figure}

We  have thus treated $v_\chi$ and $v_\xi$ as phenomenogical inputs, with their difference 
 not exceeding 30\% in our randomly generated points. In a similar fashion, the  parameters
in all scalar mass-squared matrices have been subjected to random variation, without differing by more
than 30\% with respect to values yielding $v_\chi = v_\xi$. After  diagonalizing the mass matrices, the couplings of the physical scalar states have been calculated using Feynrules\cite{feyn}. It has been checked that replacing
30\% by 50\% in the random number generation criteria do not make any qualitative change
in the final results presented below. Apart from these, the general guidelines followed in obtaining
the scatter plots (Figures \ref{4} - \ref{9})  are similar to those used in the previous section.

The quantity $s_H$, which in the previous cases uniquely parameterised the triplet contribution
to the `effective vev' breaking the electroweak symmetry, is not similarly applicable when $v_\chi \ne v_\xi$.
We instead make use of the quantity

\begin{equation}
\frac{v_{triplet}}{v} = {\frac{ {\sqrt{4(v^2_\chi + v^2_\xi)}}} {v}} 
\end{equation}
where $v = \sqrt{v^2_\Phi + 4(v^2_\chi + v^2_\xi)}$ is the effective vev. Since the custodial symmetry is broken, the physical states no more constitute the irreducible representations under the custodial $SU(2)$ group. Hence all the $H_5$ states will now have non-zero component coming from the $SU(2)_{L}$ doublet $\Phi$\footnote{The mixing between the custodial 3-plet and the 5-plet will in principle also allow $H_5^{++} \rightarrow H_5^+W^+$ resulting in a new two channel($H_5^{++} \rightarrow W^+W^+$, $H_5^{++} \rightarrow H_5^+W^+$), three channel($H_5^{++} \rightarrow W^+W^+$, $H_5^{++} \rightarrow H_5^+W^+$,$H_5^{++} \rightarrow H_3^+W^+$) and even four channel($H_5^{++} \rightarrow W^+W^+$, $H_5^{++} \rightarrow H_5^+W^+$,$H_5^{++} \rightarrow H_3^+W^+$,$H_5^{++} \rightarrow H_3^+H_3^+$) decay mode. However the unitarity of such mixing implies that the constraints will not be strengthened further}. Similar effects will be seen in states belonging to the other representations of the custodial $SU(2)$.The contribution of $\phi^{0^{r}}$ in $H$ after this is denoted by $\beta_{12}$.

Fig. \ref{4} shows the allowed regions in the space spanned by $m_{H_5^{\pm\pm}}$ and $\frac{v_{triplet}}{v}$,
with 4(a), 4(b) and 4(c) corresponding to the  single, double and triple channel decays of the doubly charged scalar.  The different coupling strength modification factors of the custodial singlet $H$ are presented in Figures \ref{4}-\ref{7}.  It is  to be noted in this context that the $WW$ and $ZZ$ coupling strengths for the state H are
liable to differ for $v_\chi \ne v_\xi$, a feature that affects the modification of interaction strengths in comparison with those for the SM Higgs. The allowed branching ratios for 
$H\rightarrow \gamma\gamma, Z\gamma$ are shown in Fig. \ref{8}. Fig. \ref{9} demonstrates the allowed
points in the space spanned by the values of the $\kappa_{H^{+}_5 W^{-} Z}$ and $\kappa_{H^{+}_3 W^{-} Z}$, following
the expressions given in section 2. 

The salient points noticed in Figures \ref{4}-\ref{9} are as follows:

\begin{itemize}
\item The maximum value of ${\frac{v_{triplet}}{v} }$ remains comparable to that of $s_H$ in the custodial
symmetry preserving case, when one considers situations allowing single-and double channel decays of the doubly charged Higgs. For three allowed channels, however, the constraints are relatively relaxed here, as the
allowed inequality of the real and complex triplet vevs opens up some additional phase space,  allowing $H^{++}_5 \rightarrow H^+_3 H^+_3$ and thus eating into the branching ratio 
for $H^{++}_5 \rightarrow W^+ W^+$, the primary search channel.

\item $H$, the custodial singlet neutral scalar, can now have a larger number of allowed points
with substantial  doublet content, especially for cases with double and triple channel decays allowed, as seen from
Fig. \ref{5}-\ref{7}. 

\item As compared to the case with exact custodial symmetry, one can have larger mass splitting
allowed between two custodial multiplets. It is thus possible to have allowed regions in the parameter space with the state $H$ as high as a TeV, with the erstwhile 3-and 5-plet
state lying as low as 500 GeV.

\item Here too, there are allowed regions in all three cases  with enhanced branching ratios in the $\gamma\gamma$ and
$Z\gamma$ channels for $H$-decay. In general, for the parameter points where such enhancement occurs, the deciding factor turns out to be not only the vev split but also their absolute values, a fact that may not be visible in the scatter plots.

\item Because of the added freedom of differencing $v_\chi$ and $v_\xi$, the maximum permissible strength of the $H^+_5 W^- Z$ interaction can be stronger than that  with $v_\chi = v_\xi$ over a non-negligible region of the parameter space. However, such interaction is found to be appreciably enhanced with respect to the case with $v_\chi = v_\xi$, when $H_5^{++}$ exclusively decays to $W^+ W^+$ that is to say in the single channel case. This effect is less pronounced for the two and three channel cases where the limit is already relaxed for $v_\chi = v_\xi$.
\item In practically  all cases Fig. \ref{4} onwards, the allowed parameter regions answering to three decay channels for $H_5^{++}$ are rather restricted, largely due to the interconnected nature of parameters and the proliferation of conditions to satisfy. For the same reason, the `allowed' or `disallowed' label of any point in a plot may get altered upon minute variation in parameter values. For example  such status may be effected $s_H$ altered in the second/third place after decimal.The broadly allowed regions are nonetheless represented faithfully.
\item The possibility of mixture of the 5-and 3-plets allows tree-level  $H^+_3 W^- Z$ interaction
in this case. Such coupling is less abundant for the triple channel $H_5^{++}$ decay, since this mass hierarchy allow very few points which simultaneously pass through the checks imposed by HiggsBounds and HiggsSignals. 
 \end{itemize}

\section{Summary and conclusion}
We have made an extensive analysis of the constraints on the parameter space
of the GM scenario, based on existing data. These includes collider data
(including VBF/Drell-Yan data at the LHC with integrated luminosity
of 137/139~fb$^{-1}$),
those on the SM-like 125-GeV scalar, indirect limits including those
from rare heavy flavour decays, and also all theoretical guidelines such
as vacuum stability and unitarity. Searches for the doubly-charged scalar
constitute the most spectacular way of probing such a scenario. We have gone
beyond the usually adopted idea that the $W^+ W^+$ decay channel is the
only significant one when it comes to situations with substantial triplet
contributions to the weak gauge boson masses. Thus we have carried detailed
scans of the parameter space where the $W^+ H^+_3$ and $H^+_3  H^+_3$
decay modes also open up, thus eating into the branching ratio share of the $W^+ W^+$ mode.

It is found that, with all possibilities and constraints included, the the upper limit
on the value of $s_H$, a measure of the triplet contribution to the $W$-and
$Z$-masses, goes up to about $0.4$ in with mass around $ 500$ GeV,
in contrast to studies based on earlier data where similar values are attainable
for masses close to a TeV only, and with constraints coming from data
with lower luminosity. We also note that the constraints from unitarity
tend to suppress the maximum value of $s_H$ for large ($\ge 1.5$ TeV)
$m_{H_5^{\pm\pm}}$, thanks to the relations
among physical masses and quartic couplings.
The $H^+ W^-Z$ couplings correspondingly have the scope of enhancement
when two-and three-channel decays of the doubly charged scalar are allowed.
We also note that, in such situations, the $\gamma\gamma$ and $Z\gamma$
branching ratios for the custodial singlet scalar $H$ can sometimes be much
higher than that of the SM-like scalar state.

While the above conclusions apply to the case with the custodial SU(2) symmetry  intact, we extend our study to situations where the real and complex triplets have unequal vev's. In a phenomenological approach to parameter
scans, we have demonstrated that  a larger number of parameter points can become allowed,
subject to all constraints, and the features outlined above are more widely visible.
In addition the charged scalar $H^+_3$, too, has $W^-Z$ interaction here. Thus the GM scenario still admits of interesting phenomenology, subject to the various
constraints that tend to tie it up. 

\section{Acknowledgment} The authors thank Utpal Sarkar, Tousik Samui and Ritesh K Singh for helpful discussion. The work of R.G has been supported by a fellowship awarded by University Grants Commission, India.



\end{document}